\documentclass[letterpaper,prc,twocolumn,showpacs,floatfix,nofootinbib,preprintnumbers,superscriptaddress,amsmath,amssymb]{revtex4-1}
\usepackage{epsfig}
\usepackage{graphicx}
\usepackage{amssymb}
\usepackage{color}
\usepackage{amsmath}
\usepackage[colorlinks=true]{hyperref}

\usepackage{isotopes}
\newcommand{\qvec}{\mathbf{q}}
\newcommand{\rvec}{\mathbf{r}}

\newcommand{\HFB}{\textsc{hfb}}
\newcommand{\TDGCM}{\textsc{tdgcm}}
\newcommand{\GOA}{\textsc{goa}}
\newcommand{\FELIX}{F\textsc{elix}}
\newcommand{\GCM}{\textsc{gcm}}
\newcommand{\ATDHF}{\textsc{atdhf}}
\newcommand{\ZPE}{\textsc{zpe}}
\newcommand{\DONES}{\textsc{d1s}}
\newcommand{\DONEN}{\textsc{d1n}}
\newcommand{\DONEM}{\textsc{d1m}}

\newcommand{\HO}{\textsc{ho}}
\newcommand{\PES}{\textsc{pes}}
\newcommand{\EDF}{\textsc{edf}}

\graphicspath{{figures/}}

\begin{document}

\title{From Asymmetric to Symmetric Fission in the Fermium Isotopes within the Time-Dependent GCM Formalism}
  
\author{D.~Regnier} 
\email{regnier@ipno.in2p3.fr}
\affiliation{Institut de Physique Nucl\'eaire, IN2P3-CNRS, Universit\'e Paris-Sud, Universit\'e Paris-Saclay, F-91406 Orsay Cedex, France}

\author{N.~Dubray}
\email{noel.dubray@cea.fr}
\affiliation{CEA,DAM,DIF, 91297 Arpajon, France}

\author{N.~Schunck}
\email{schunck1@llnl.gov}
\affiliation{Nuclear and Chemical Science Division, LLNL, Livermore, CA 94551, USA}
  
\date{\today}

\begin{abstract} 
\begin{description}
\item[Background]
Predicting the properties of neutron-rich nuclei far from the valley of stability is one of the major challenges of modern nuclear theory. In heavy and superheavy nuclei, a difference of only a few neutrons is sufficient to change the dominant fission mode. A theoretical approach capable of predicting such rapid transitions for neutron-rich systems would be a valuable tool to better understand $r$-process nucleosynthesis or the decay of super-heavy elements.
\item[Purpose]
In this work, we investigate for the first time the transition from asymmetric to symmetric fission through the calculation of primary fission yields with the time-dependent generator coordinate method ({\TDGCM}). We choose here the transition in neutron-rich Fermium isotopes, which was the first to be observed experimentally in the late seventies and is often used as a benchmark for theoretical studies.
\item[Methods]
We compute the primary fission fragment mass and charge yields for \Fm254,\Fm256 and \Fm258 from the {\TDGCM} under the Gaussian overlap approximation. The static part of the calculation (generation of a potential energy surface) consists in a series of constrained Hartree-Fock-Bogoliubov calculations based on the {\DONES}, {\DONEM} or {\DONEN} parameterizations of the Gogny effective interaction in a two-center harmonic oscillator basis. The 2-dimensional dynamics in the collective space spanned by the quadrupole and octupole moments $(\hat{Q}_{20}, \hat{Q}_{30})$ is then computed with the finite element solver {\FELIX}-2.0.
\item[Results]
The available experimental data and the {\TDGCM} post-dictions are consistent and agree especially on the position in the Fermium isotopic chain at which the transition occurs. In addition, the {\TDGCM} predicts two distinct asymmetric modes for the fission of \Fm254.
\item[Conclusions]
Thanks to its intrinsic accounting of shell effects and to its ability to describe the dynamics of the system up to configurations close to scission, the {\TDGCM} is able to describe qualitatively the fission yield transition in the neutron-rich Fermium isotopes. This makes it a promising tool to study the evolution of the fission yields far from the valley of stability. The main limitation of the method lies in the presence of discontinuities in the 2-dimensional manifold of generator states.
\end{description}
\end{abstract}

\pacs{25.85.Ec, 24.10.Cn, 21.60.Jz, 21.10.Gv}
  
\keywords{nuclear fission, generator coordinate method, Fermium}
\maketitle

\section{Introduction}

One of the goals of nuclear theory is to provide models that not only reproduce a large set of available experimental data in the neighborhood of the valley of stability, but also have predictive power when computing properties of nuclei far from this region. On-going efforts to better understand the $r$-process of nucleosynthesis or the decay of super-heavy elements give a particular stake to the neutron-rich part of the nuclear chart. In the special case of the fission process, it was shown in Ref.~\cite{goriely_first_2009} that the discrepancies between models used to determine the fission fragment yields of the neutron-rich systems involved in the $r$-process may significantly impact the predictions of the abundances in the region of rare earth peak. This is an incentive to develop a theoretical framework capable of predicting the fission properties for a wide range of neutron-rich nuclei.

At the same time, it is also known that the properties of fission fragments may vary drastically with the number of neutrons and protons of the fissioning system. Historically, a series of experiments conducted in the 70-80's showed that adding only a few neutrons to \Fm254 could totally change the dominant low-energy fission mode. In these experiments, the post-neutron emission fragments were characterized either by radio-chemistry or directly by measuring their kinetic energy. 
For \Fm254, the mass yields were obtained from spontaneous fission~\cite{harbour_mass_1973,gindler_distribution_1977} and clearly showed a mostly asymmetric behavior.
When adding a few neutrons, this asymmetric feature is less sharp. The fission yields of \Fm256 both from the neutron-induced channel~\cite{ragaini_symmetric_1974,flynn_mass_1975} and the spontaneous fission channel~\cite{flynn_distribution_1972,bemis_mass_1977} all exhibit a mostly asymmetric behavior but the group of Ragani {\em et al.} detected in addition the presence of an appreciable symmetric component. For the spontaneous fission of \Fm257, two papers by Balagna {\em et al.}~\cite{Balagna_mass_1971} and John {\em et al.}~\cite{john_symmetric_1971} reported contradictory results on the dominant fission mode.
Finally, symmetric fission clearly dominates in \Fm258 as reported in three papers covering both the neutron-induced and spontaneous fission~\cite{flynn_distribution_1975,hulet_spontaneous_1980,hulet_spontaneous_1989}. The group of Hulet {\em et al.} even probed the fission of \Fm259 produced by {\Fm257}(t,p) and found out mostly symmetric yields. 
Later on, several experiments relying on inverse kinematic beams~\cite{schmidt_relativistic_2000,martin_studies_2015} highlighted many similar transitions both in the neutron-rich and neutron-deficient sides of the valley of stability. A common feature is that the transition often occurs within a range of just a few nucleons. 

Understanding and reproducing these sharp transitions presents a real challenge for nuclear theory and different kinds of approaches have been proposed to tackle this issue. A common starting point is often the computation of the potential energy surface for the fissioning system as a function of a small set of collective degrees of freedom.
In 1980, Lustig {\em et al.} were the first to study the asymmetric/symmetric transition of mass yields in Fermiums. They adopted a purely static picture and computed the energy of the deformed nucleus within a macroscopic-microscopic model~\cite{lustig_transitions_1980}. Later on, a similar work performed by Cwiok {\em et al.}~\cite{cwiok_two_1989} in a five-dimensional deformation space revealed the existence of an elongated and a compact fission mode for \Fm258.
More recently, studies of static deformation properties of Fermium isotopes were also performed within a self-consistent mean-field framework based on Gogny, Skyrme and covariant energy density functionals ({\EDF})~\cite{warda_self-consistent_2002,warda_spontaneous_2006,bonneau_fission_2006,dubray_structure_2008,staszczak_microscopic_2009,zhao_multidimensionally-constrained_2016}. All these papers emphasized the multi-modal character of the fission of Fermium isotopes near $A=256$ and highlighted the presence of three major modes: symmetric compact, symmetric elongated, and asymmetric. 
Although these static approaches pinpointed the major fission modes that are energetically favored in low-energy fission, they did not provide information about the actual probability to populate each of these modes.

One way to predict fission yields without an explicit treatment of nuclear dynamics is to assume that static nuclear configurations close to scission are populated statistically during the fission process. Such scission-point models have been applied with different choices for the deformed nuclear configurations~\cite{schmidt_review_2018, pasca_charge_2018,ichikawa_origin_2009,lemaitre_new_2015}. 
These models were able to reproduce the main features of the symmetric/asymmetric transitions of the fission yields for instance in the Thorium and Fermium isotopic chains. 
However, one of the major limitations of scission point models is the somewhat arbitrary definition of the ensemble of scission configurations which are thermally populated. One should also keep in mind that they ignore any possible ``memory effect'' of the nucleus as it travels through the potential energy landscape.

%
Another class of approaches to determine fission yields involve using static nuclear properties as inputs to the explicit modeling of nuclear dynamics. Following this idea, Asano {\em et al.} performed Langevin calculations in three-dimensional collective spaces~\cite{asano_dynamical_2004}. This represented the first theoretical attempt to obtain the yields of \Fm256,\Fm258,\Fm264 through the proper simulation of the time-evolution of the system. However, the calculation failed to reproduce the observed transition from asymmetric to symmetric mass yields between \Fm256 and \Fm258.

A fully quantum mechanical alternative to describe nuclear dynamics is the time-dependent generator coordinate method ({\TDGCM}) with the Gaussian overlap approximation ({\GOA}). Goutte \textit{et al.} used this framework for the first time in 2005 to compute fission yields~\cite{goutte_microscopic_2005}. Since then, it has been successfully applied to several fissioning systems in the actinide region~\cite{younes_fragment_2012,regnier_fission_2016,zdeb_fission_2017} including proton-rich Thorium isotopes~\cite{tao_microscopic_2017}. However, the reliability of this method in the neutron-rich sector of the nuclear chart and its ability to predict rapid structural changes in fission yields is yet to be established.

The goal of this paper is to investigate the robustness of the {\TDGCM}+{\GOA} approach in reproducing the symmetric/asymmetric yield transition in neutron-rich Fermium isotopes. In particular, we will examine in details the dependence of the results on the various inputs to the calculations: parametrization of the energy density functional, initial conditions, form of the collective inertia tensor, and definition of scission configurations.

In Sec.~\ref{sec:theory} we briefly recall the formal and numerical methods used to compute fission yields within the {\TDGCM}+{\GOA}. The Sec.~\ref{sec:results} is devoted to the discussion of the static properties obtained for the Fermium chain and to the comparison between the computed fission yields and associated experimental data. In Sec.~\ref{sec:discussion} we focus on the reliability of this approach by testing the sensitivity of our results to various input ingredients. We also analyze the impact of discontinuities in the 2-dimensional manifold of generator states.

%
%

\section{Methodology}
\label{sec:theory}
A comprehensive presentation of the {\TDGCM}+{\GOA} theory and of our implementation of it can be found in Ref.~\cite{regnier_fission_2016}. In this section we only summarize the necessary ingredients of the method and refer the reader to our previous work for further details. 

\subsection{Theoretical Framework}

In the {\TDGCM} approach, the evolution of the many-body quantum state $|\Psi(t)\rangle$ describing the fissioning system is determined by a variational approximation of the many-body dynamics. At any time, the many-body wave function takes the form of a continuous and linear superposition of constrained {\HFB} states parametrized by a set of collective coordinates $\qvec$,
\begin{equation}
\label{eq:gcmApprox}
|\Psi(t)\rangle \equiv \int_{\qvec} f(\qvec, t) |\Phi_{\qvec} \rangle \, \text{d}\qvec.
\end{equation}
Instead of solving the non-local Hill-Wheeler equation resulting from the application of the time-dependent variational principle, we invoke in addition the Gaussian overlap approximation ({\GOA})~\cite{brink_generator-coordinate_1968,reinhard_generator-coordinate_1987,krappe_theory_2012}. This standard scheme reduces the problem to a local Schr\"odinger-like equation,
\begin{equation}
\label{eq:tdgcmgoa}
i\hbar \frac{\partial g(\qvec,t)}{\partial t} = \hat{H}_\text{coll}(\qvec) \, g(\qvec,t).
\end{equation}
The complex function $g(\qvec,t)$ is the unknown of the equation. It is related to the weight function $f(\qvec, t)$ appearing in (\ref{eq:gcmApprox}) and contains all the information about the dynamics of the system. The collective Hamiltonian $\hat{H}_\text{coll}(\qvec)$ is a local linear operator acting on $g(\qvec,t)$,
\begin{equation}
\label{eq:Hcoll}
\hat{H}_\text{coll}(\qvec) \equiv
-\frac{\hbar ^2}{2\gamma^{1/2}(\qvec)} \sum_{ij} \frac{\partial}{\partial q_i} \gamma^{1/2}(\qvec) B_{ij}(\qvec) \frac{\partial}{\partial q_j}  +  V(\qvec).
\end{equation}
This operator contains a collective kinetic part characterized by the inertia tensor $\mathsf{B}(\qvec) \equiv B_{ij}(\qvec)$ and a potential term $V(\qvec)$. In a generalized version of the {\GOA}~\cite{kamlah_derivation_1973,onishi_local_1975,gozdz_extended_1985}, it also involves a real and positive metric $\gamma(\qvec)$. Taking into account this metric leads to a better reproduction of the exact overlaps with Gaussian functions by letting the width of the Gaussian kernels explicitly depend on the position in the collective space.
The locality of the collective Hamiltonian implies a continuity equation for the square modulus of the collective wave function $|g(\qvec,t)|^2$,
\begin{equation}
\label{eq:continuity}
\frac{\partial}{\partial t} |g(\qvec, t)|^2 \gamma^{1/2}(\qvec) = -\nabla \cdot \mathbf{J}(\qvec, t),
\end{equation}
where $\mathbf{J}(\qvec, t)$ is the collective current defined from $g(\qvec,t)$.

To compute the fission yields from the solution of Eq.~\eqref{eq:tdgcmgoa}, we define a frontier line that marks the limit between (i) an inner domain of the collective space where we still have a compound nucleus and (ii) an outer domain containing eventually all the split configurations.
Within this picture, each infinitesimal element of the frontier line corresponds to the entrance point of one possible output channel of the fission reaction with a given mass and charge for the two primary fragments.
Ideally, the frontier should be chosen in such a way that output channels are completely decoupled from one another. In this situation, the collective dynamics in the inner domain would simulate the evolution up to configurations where the two fragments could not exchange particles any more. The quantum probability to measure a mass split $A_H/A_L$ would then be given by the projection of the final {\GCM} state over all output channels leading to this mass split. This is nothing but the integral of $|g(\qvec, t)|^2\gamma^{1/2}(\qvec)$ over a set of outer collective areas, each associated with one output channel. Leveraging the continuity equation Eq.~\eqref{eq:continuity}, it can be recast into a sum of time-integrated flux of probability $F(\xi,t)$ to cross an infinitesimal element $\xi$ of the frontier.
\begin{equation}
F(\xi,t) = \int_{t=0}^{t} dt \int_{\qvec\in\xi} \mathbf{J}(\qvec,t)\cdot d\mathbf{S}.
\label{eq:fluxDef}
\end{equation}
For the fragmentation $A_H/A_L$, the sum runs over all elements $\xi$ in which the {\HFB} states have $A_L/A_H$ particles in the light/heavy fragment.
In practice, the choice of the frontier is subject to several constraints discussed in Sec.~\ref{sec:front} and \ref{subsec:discont}. In our calculations, the configurations at the frontier are often characterized by a non-negligible interaction energy between the pre-fragments~\cite{younes_nuclear_2011}. This means that the {\HFB} states on the frontier do not yet fully belong to one or the other of the output channels. In other words, a realistic evolution of such a state may lead to several mass splits close to our averaged estimate at the frontier. 
To take this into account, several prescriptions have been proposed in the literature such as convoluting the raw yields with a Gaussian~\cite{younes_fragment_2012} or using a more sophisticated random neck rupture model~\cite{zdeb_fission_2017}. In this work, we retain a simple prescription and adopt a Gaussian convolution with a constant width. Doing so, we introduce the width of the Gaussian used in the convolution as a necessary arbitrary parameter.
All final yields are normalized to 200\%.

\subsection{Determination of the GCM+GOA Collective Hamiltonian}
\label{sec:genpes}

The first step to build the collective Hamiltonian consists in building the manifold of generator states. In practice, it implies performing a series of {\HFB} calculations for the compound nucleus with constraints on the expectation value of the two collective coordinates $\hat{Q}_{20}$ and $\hat{Q}_{30}$, which are here defined with the same conventions as in Ref.~\cite{regnier_fission_2016}. 
We computed each point in the regular grid spanning $[0,450]\times[0,100]$ (in barn units) with the mesh steps $h_{20}=2$ b, and $h_{30}=1$ b$^{3/2}$.
Each {\HFB} calculation is performed by an iterative solver relying on a two-center {\HO} basis to discretize the single particle wave functions. The parameters of this basis are optimized at each deformation point using a new method based on Gaussian processes. This new method, which will be described in details in a future paper, allowed us to speed up the basis parameter optimization procedure by a factor 5, compared to the previous numerical procedure. The {\HFB} calculations have been performed with the {\DONES}, {\DONEN} and {\DONEM} parameterizations of the Gogny effective interaction for each of the three Fermium isotopes.

It is well known that generating a potential energy surface which minimizes the total binding energy (as is the case using self-consistent methods) may lead to some issues related to the imperfect nature of the minimization (local minima) and to the underestimation of some barrier heights (restricted collective space) as discussed in Ref.~\cite{dubray_numerical_2012}. To fully avoid the issue of spurious local minima, a special retro-propagation scheme is used, which ensures that all {\HFB} solutions of the potential energy surface are global minima. 

From the ensemble of {\HFB} solutions, the last step is to determine the collective fields involved in Eq.~\ref{eq:Hcoll}. In the {\GCM}+{\GOA} formalism, the inertia tensor is related to the second order derivatives of the reduced Hamiltonian kernel with respect to the collective coordinates. In this work, the {\GCM} inertia and metric are calculated at the perturbative cranking approximation; see \cite{schunck2016} for details.
The potential term provided by the {\GCM}+{\GOA} approach contains the total {\HFB} energy of the constrained state corrected by a vibrational zero-point energy associated with our collective degrees of freedom. The formula used to compute the fields can be found in Eqs. (9)-(15) of Ref.~\cite{regnier_fission_2016}.

\subsection{Solution to the Collective Schr\"odinger Equation}
\label{sec:dynamics}

After the calculation of the static properties of the system, we numerically solve the collective Schr\"odinger-like equation, Eq.~\eqref{eq:tdgcmgoa}, with the version 2.0 of the code {\FELIX}~\cite{regnier_felix-1.0:_2016,regnier_felix-2.0:_2018}. 
We simulate the collective evolution in a symmetric domain for the octupole moment with absorption boundary conditions to avoid spurious reflections. To get the best numerical efficiency, the problem is not discretized on the regular mesh used to compute the static properties but on a refined and adapted finite element mesh. The technical details on the simulation domain, boundary conditions, generation of the mesh and spectral elements basis are reported in App.~\ref{ap:setup_param}.

The initial state is built as prescribed in~\cite{regnier_fission_2016} as a superposition of collective eigen-modes in an extrapolated first potential well. The weights of this mixture have a Gaussian shape as a function of the eigen-energies of the modes. The width of the Gaussian is fixed to $\sigma_i=0.5$ MeV and its first moment is chosen so that the initial energy lies 1 MeV above the first potential barrier. 
This choice for the initial collective state distribution simulates the low-energy induced fission while it allows a significant part of the wave packet to escape from the first potential well and evolve toward scission. The sensitivity of our results to the parameter $\sigma_i$ and the initial energy is discussed in Sec.~\ref{sec:ei}.

Starting from this initial condition, the evolution equation is integrated in time using a Krylov approximation scheme for the exponential propagator. We use a dimension-10 Krylov space along with a time step $dt=2\times 10^{-4}$ zs ($10^{-21}$s). The propagation runs up to a time of 20 zs, after which the fission yields are stable with time. According to our previous benchmark on \Fm256, the absolute numerical convergence of the resulting mass yields (normalized to 200\%) is expected to be of the order of 0.06\%.

\subsection{Extraction of Fission Mass Distributions}
\label{subsubsec:yields}

The frontier used to compute the fission yields is defined by the isoline $Q_{\rm N}=7.5$ of the neck operator~\cite{younes_fragment_2012}
\begin{equation}
 \hat{Q}_N = \operatorname{exp}\left( - \frac{ (z - z_N)^2}{a_N^2} \right),
\end{equation}
with $a_N = 1$ fm, $z$ the coordinate along the main axis of the system and $z_N$ is the position of the neck.
This line is chosen as one of the lowest-value neck isoline that lies above the fission/fusion valley crossing. This choice is discussed in more details in Sec.~\ref{sec:front}. In practice, the isoline is discretized as a succession of square cells edges four times smaller than the finite element mesh cell edges.
The raw yields extracted from the time-integrated flux through the frontier are convoluted with a normal distribution as already done in Ref.~\cite{regnier_fission_2016}. Such a convolution implements our lack of knowledge on the exact number of particles in the fragments due to several features that we briefly recall below:
\begin{itemize}
\item 
After solving the {\TDGCM}+{\GOA} evolution, a proper quantum estimation of the number of particles in each fragment would require first disentangling the two fragments \cite{younes_nuclear_2011,schunck2014}, and then projecting on states with a good particle number, e.g. as in \cite{scamps_superfluid_2015}. Since in this work, we only estimate particle numbers based on the integration of the one-body density, we therefore miss some of the quantum fluctuations.
\item 
By construction, the {\HFB} theory used to determine the generator states breaks the symmetry associated with the total number of particles in the fissioning system. This implies that at the frontier where the yields are computed, the total wave function of the fissioning system is the superposition of wave functions with different numbers of particles. Once again, a better approach would involve projecting this wave function on good particle number and extracting the characteristics of the fragments from the projected density.
\item 
For the nuclear configurations at the frontier, the nuclear interaction between the fragments can easily be of the order of dozens of MeV; see, e.g., estimates in \cite{schunck2014}. This implies that several nucleons could be exchanged between the two prefragments. Each configuration at the frontier therefore contributes to several neighboring fragmentations.
\item 
Finally, the experimental fission yields that we have used in this study were measured with a detector resolution of 4-5 mass units (full width at half maximum). This corresponds to a convolution of the raw yields with a normal distribution parameterized with a width $\sigma \simeq 2$. A similar convolution of the theoretical results should normally be applied in order to make consistent comparisons.
\end{itemize}

Addressing these limitations goes beyond the scope of this article. At the moment, we therefore make the pragmatic choice of taking effectively into account these effects by reducing the resolution of our predictions. To do so we convolute the raw fission yields with a Gaussian of width $\sigma=4$ mass units.
The choice of this parameter can be justified based on various physical arguments. Indeed, we have $Q_{\rm N} = 7.5$ for configurations at the frontier, which means that roughly 8 particles are located in a plane within $\pm 1$ fm around the neck  position. If the radial total density is constant in this region, and if we assume a random rupture of the neck with a normal probability distribution $P(x_{\text{neck}} + \delta x)$ for the split to happen at $x_{\text{neck}} + \delta x$, then we obtain a spreading with the same width ($\sigma\simeq 4$ mass units). In others words making such a convolution on the fragment mass is equivalent to considering that the neck is randomly cut with a probability following a normal distribution of width 1 fm.
While this reasoning provides a qualitative motivation for the choice of the convolution width, it should be clear that the precise quantitative value of the convolution width $\sigma$ is still arbitrary.
Hopefully, our previous study~\cite{regnier_fission_2016} shows that changing this value does not impact significantly the main characteristics of the fission modes.

\section{Results}
\label{sec:results}

In this section we present the static and dynamic properties of $^{254,256,258}$Fm obtained within the {\TDGCM}+{\GOA} approach. 

%
\subsection{Main Static Properties}

For each nucleus, we first computed the generator states with the {\DONES} parameterization of the Gogny effective interaction. The fitting process of this parameterization includes information on the fission barrier of \Pu240, which makes it a reference effective interaction for fission studies in general. Unless specified otherwise, the calculations presented in this section are based on Gogny {\DONES}.

\subsubsection{Global Topology}
\begin{figure}[!ht]
\includegraphics[width=0.5\textwidth]{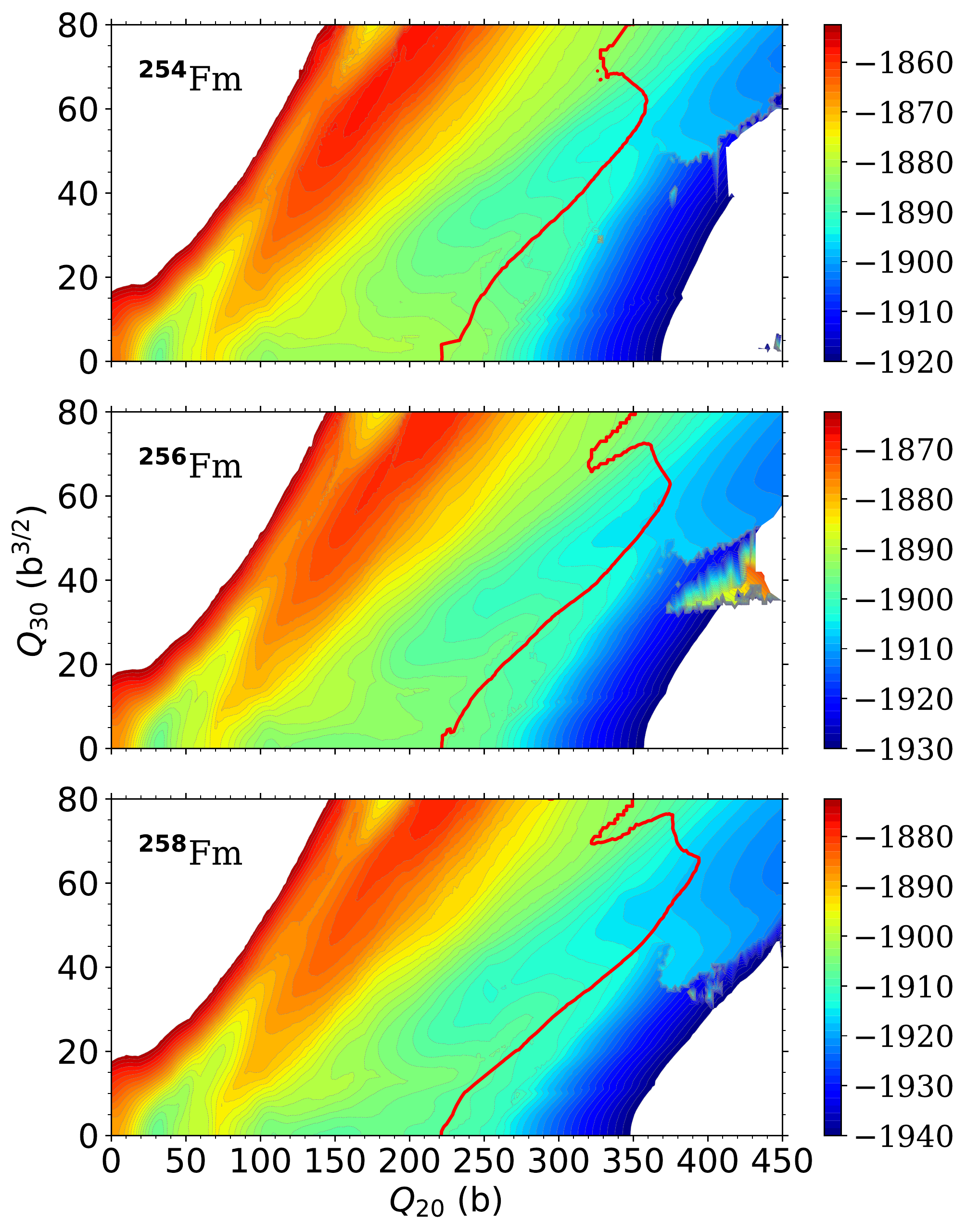}
\caption{Potential energy surfaces of the 254, 256 and 258 Fermium isotopes determined from the {\DONES} Gogny energy density functional. The potential corresponds to the {\HFB} energy corrected from the {\GCM} zero-point energy. The color scale is shifted by 10 MeV between consecutive plots. The red continuous line represents the isoline $Q_{\rm N} = 7.5$ of the neck operator.}
\label{fig:fmtrans_pes}
\end{figure}

The figure~\ref{fig:fmtrans_pes} shows the potential energy landscape obtained for the three nuclei under study. Note that the potential includes the {\GCM} zero-point energy.
%
%
The overall topology of theses potential energy surfaces ({\PES}) is very similar for the three nuclei. The energy minimum in the first potential well is characterized by $Q_{20}\approx 30$ b and $Q_{30}\approx 0$ b$^{3/2}$. This is typical of the actinide region. Going toward more elongated shapes, there is a first potential barrier whose height depends on the specific nucleus. In our 2-dimensional collective space, two main fission modes are clearly visible. The first one is a rather broad valley (in the $Q_{30}$ direction) leading to asymmetric fragmentations. It reaches neck values $Q_{\rm N}=7.5$ at large elongations $Q_{20}\in [350,400]$ b and corresponds to what is called the asymmetric elongated fission mode.
The second fission mode is a tiny valley that follows symmetric configurations and reaches the same neck value at much lower elongation ($Q_{20}\simeq 220$ b). Beyond this line, a rapid change in the energy slope happens around $Q_{20}\simeq 260$ b. The collective potential energy decreases rapidly and the expectation value of the neck operator also vanishes suddenly. It corresponds to the symmetric compact fission mode discussed in Ref.~\cite{bonneau_fission_2006,staszczak_microscopic_2009}. For \Fm258, a third symmetric elongated mode has also been described in these previous papers but it is not visible here. This is because our 2-dimensional {\PES} can only show the lowest energy modes in a given range of $Q_{20},Q_{30}$ whereas the symmetric elongated and compact fission modes span the same range for these collective variables. As shown in Ref.~\cite{staszczak_microscopic_2009}, introducing an additional dimension through the $Q_{40}$ collective coordinate would enable us to capture both symmetric modes. Note that calculations in 3-dimensional collective spaces \cite{bonneau_fission_2006,staszczak_microscopic_2009} suggest that the symmetric elongated mode lies quite higher in energy than the symmetric compact mode. Therefore, its contribution to the formation of the symmetric peak in fission fragment distributions should not be significant.

\begin{table}[!htb]
\begin{tabular}{ccccc}
\hline
         & $V_{\text{min}}$ & $E_{0,\text{GCM}}$  & B$_I$ & B$_{I, \text{GCM}}$ \\
\hline
\Fm254 & -1886.2 & -1883.0  & 13.3 & 10.2    \\
\Fm256 & -1896.6 & -1893.6  & 12.4 & 9.4     \\
\Fm258 & -1906.8 & -1903.7  & 11.8 & 8.7     \\
\hline
\end{tabular}
\caption{Characteristics of the Gogny {\DONES} potential energy surfaces for $^{254,256,258}$Fm. The minimum  of the potential ($V_{\text{min}}$) in the first well is given in MeV along with the energy of the {\GCM} ground-state ($E_{0,\text{GCM}}$). The height of the inner fission barrier (B$_I$) is in MeV relative to $V_{\text{min}}$. The quantity B$_{I, \text{GCM}}$ is the energy that should be brought to the system in its {\GCM} ground state in order to fission without tunnel effect.}
\label{tab:barriers}
\end{table}

In details, the heights of the different barriers and ridges significantly differ from one nucleus to another. We show in Table~\ref{tab:barriers} the first fission barrier heights relative to the minimum energy in the first potential well.
A quantity that has more physical relevance than the barrier is the quantity of energy that must be injected in the compound nucleus so that the fission process may happen without tunneling. In our framework this is given by the difference between the potential at the saddle point and the energy of the {\GCM} ground-state in the first potential well. We report this quantity as B$_{I,\text{GCM}}$ and show that it is lower than the 'classical' barrier by a few MeV. This "collective" barrier could be further reduced by a few MeV if axial symmetry were not imposed in our {\HFB} calculations.

\subsubsection{Competing Fission Modes}
In the {\TDGCM}+{\GOA} picture, the presence of valleys in the potential energy landscape favors the diffusion of the collective wave packet towards specific sets of configurations at scission. As discussed in the previous section, two major valleys have been found in the present calculations (see Fig.~\ref{fig:fmtrans_pes}). These two valleys are separated by a potential ridge with a shape and height that varies with the nucleus. 
This ridge is indeed quite pronounced for \Fm258 but progressively disappears as we go toward the lighter isotopes. 
%
\begin{figure}[!ht]
\includegraphics[width=0.45\textwidth]{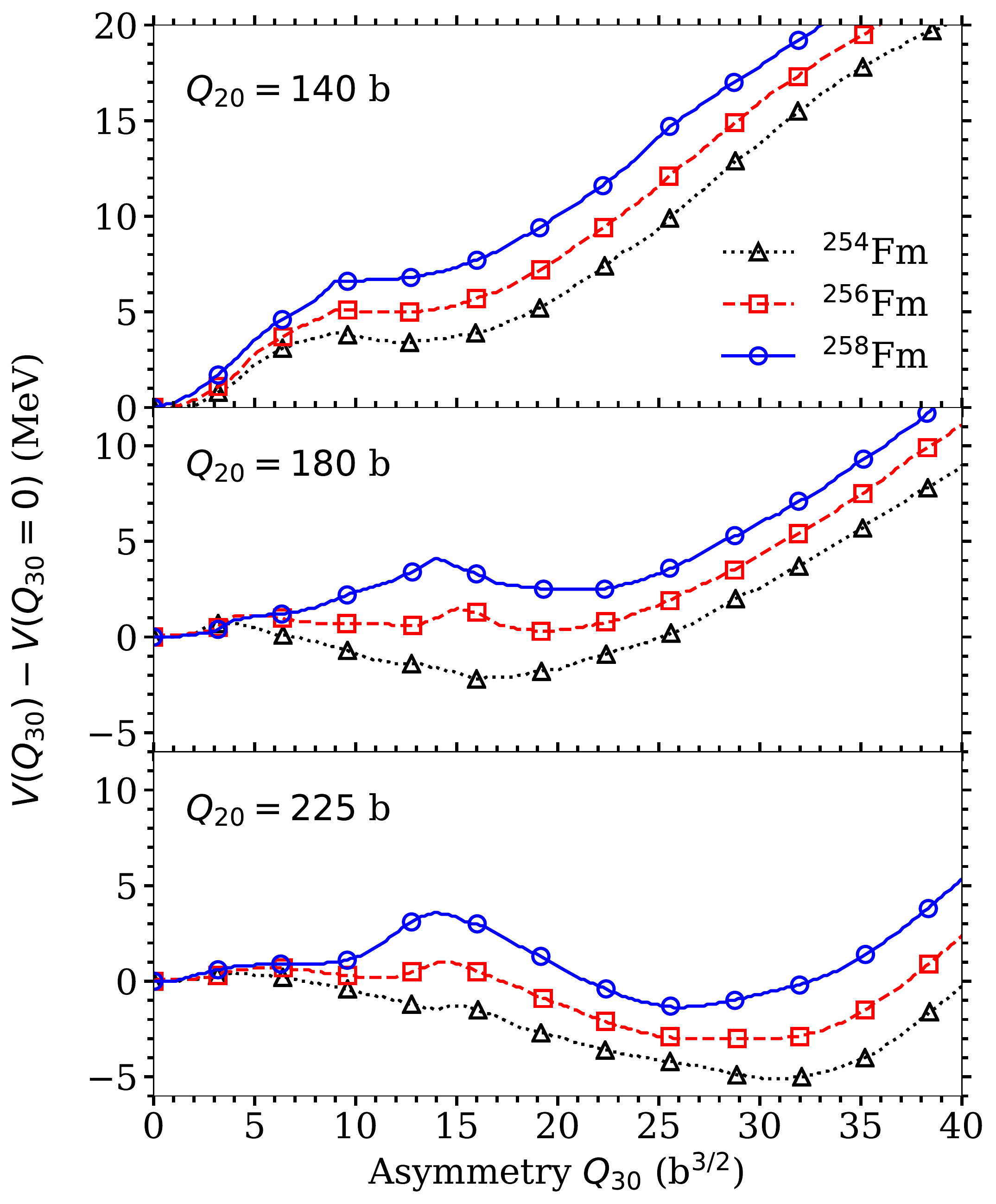}
\caption{Slice of the {\DONES} potential energy surfaces for the three Fermium isotopes at the elongations $Q_{20}=140,180,225$ b. To emphasize the difference of topology between nuclei, all the curves are shifted so that $V(Q_{30}=0) = 0$.}
\label{fig:vcut}
\end{figure}

We quantify this behavior in Fig.~\ref{fig:vcut}, which shows slices of the three {\PES} at the constant quadrupole moment values $Q_{20} = 140, 180, 225$ b. 
At $Q_{20}=140$ b, symmetric configurations are largely favored energetically in all three isotopes. Around $Q_{20}=180$ b, the symmetric path is favored in \Fm258 but, in contrast, the asymmetric mode is lower in energy for \Fm254. Since there is no significant potential barrier between the two valleys, the system can diffuse from symmetric configurations to asymmetric configurations at the mouth of the asymmetric valley. In \Fm256, the {\PES} is rather flat which provides the opportunity for a collective wave packet to spread over the two valleys populating both modes. At larger elongations $Q_{20}=225$ b, even if the asymmetric mode becomes energetically more favored in \Fm258, a ridge of 4 MeV separates it from the symmetric path and hinders the transition toward the asymmetric elongated mode. 
Such changes in the topology of the {\PES} are likely to be highly correlated with the appearance of gaps in the single particle energy spectra as a function of the collective deformations.
Although this analysis based on the static potential energy is not yet quantitative, most of the physics of the transition can already be guessed at that level.

%

\subsection{Fission Fragment Distributions}

We computed the {\TDGCM}+{\GOA} evolution of the \Fm254, \Fm256 and \Fm258 over a period of 20 zs ($10^{-21}$s).
During the propagation, the collective wave function $g(\qvec)$ escapes the first potential well to populate the available fission valleys. After crossing the frontier, it is then absorbed by the artificial imaginary term in the Hamiltonian in the absorption band. After 20 zs, 34\%, 34\%, and 28\% of the total norm crossed the frontier for the $A=254$, 256 and 258 Fermium isotopes respectively. 
During the last 1 zs, the yields are nearly stable and we have $|| \mathbf{Y}(t)- \mathbf{Y}(t_f)||_{\infty} < 0.4\%$ for the intermediate \Fm256 and $|| \mathbf{Y}(t)- \mathbf{Y}(t_f)||_{\infty} < 0.1\%$ for the others. This means that although some of the wave packet is still leaking from the first potential well, 20 zs is enough time to obtain the qualitative features of the yields.

\begin{figure}[!ht]
\includegraphics[width=0.45\textwidth]{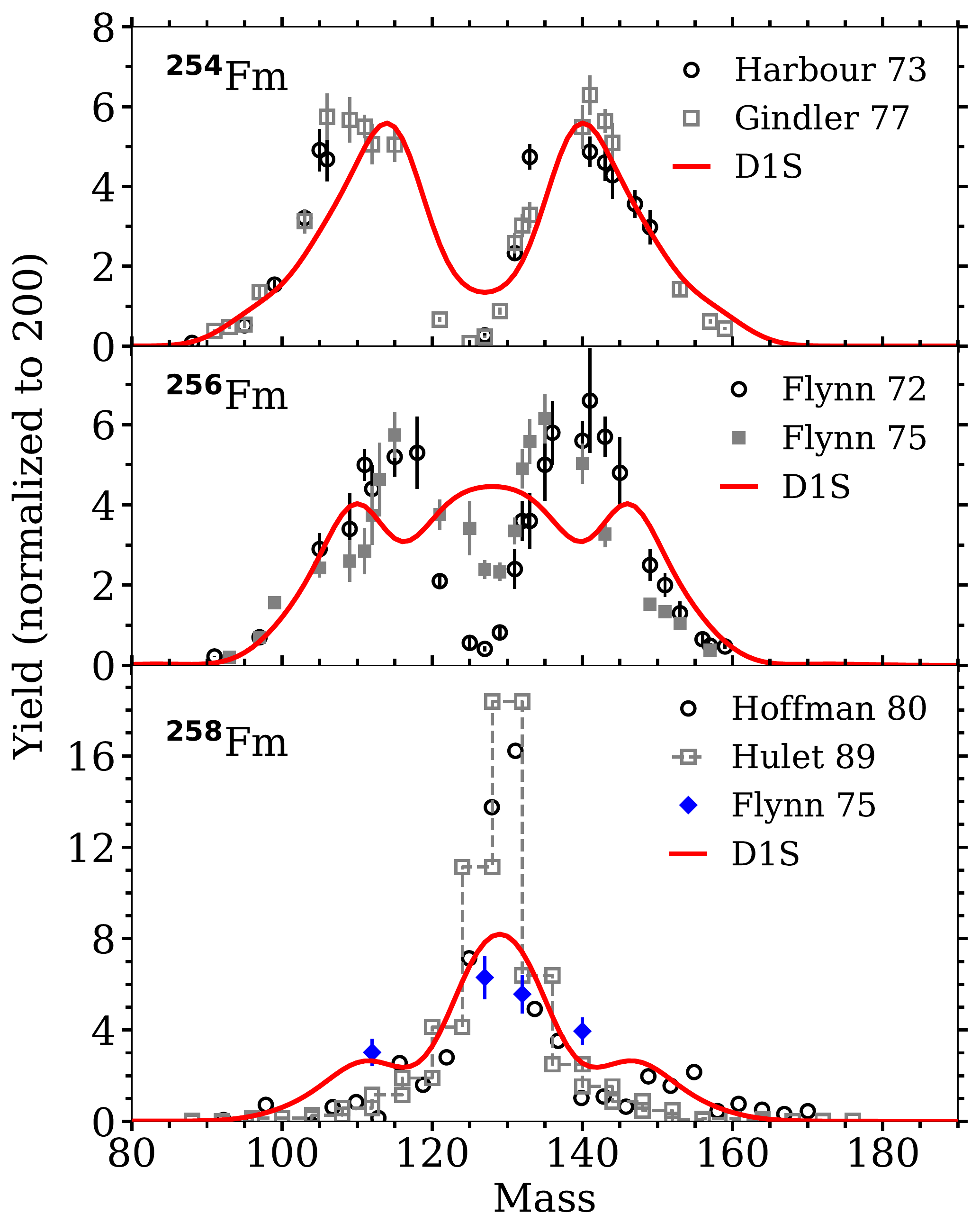}
\caption{Primary fragment mass yields obtained with the Gogny {\DONES} effective interaction and compared with various experimental data sets taken from Ref.~\cite{harbour_mass_1973,gindler_distribution_1977,flynn_distribution_1972,flynn_distribution_1975,hoffman_12.3-min_1980,hulet_spontaneous_1989}. All the yields are normalized to 200\%. The experimental data points all represent post-neutron evaporation mass yields. The open symbols stand for experimental data associated to spontaneous fission whereas full symbols are related to thermal neutron-induced fission.}
\label{fig:theovsexp}
\end{figure}

Figure~\ref{fig:theovsexp} presents the primary fission mass yields obtained for the fermium isotopic chain compared to a series of experimental data. 
%
%
When adding only four neutrons to the compound system, the behavior radically changes from a mostly asymmetric to a mostly symmetric fission. The {\TDGCM}+{\GOA} dynamics applied with the Gogny {\DONES} effective interaction successfully captures this rapid transition. The number of neutrons at which this transition is predicted, $N=156$, matches the experimental observations. On the other hand, the mass-by-mass values of the yields sometimes differ from the experimental by up to 2\% (in absolute value). In particular, the results obtained for the intermediate nucleus \Fm256 do not reproduce the double-humped shape of the experimental data.

There are several reasons why this comparison between theory and experiment must be kept at the qualitative level. First, for all experimental data the mass of the fragments is measured after the evaporation of prompt neutrons. Taking into account the neutron evaporation would shift our predictions by a few units toward lighter masses as well as bring additional structure and asymmetry between the light and heavy peaks. It could partly be responsible for the light peak of \Fm254 being roughly 7 mass units too high. The shift of the light peak depends non trivially on the fragmentation, and a first account of neutron evaporation would at least require the knowledge of the average neutron multiplicity as a function of the fragment mass. We did not apply such a correction here. 
A second important effect that also impacts the comparison with experiment is the initial energy of the fissioning system. Some of the experimental data sets are from spontaneous fission whereas others come from induced fission. In the actinide region, where fission is mostly asymmetric, adding more energy to the system is known to enhance the symmetric component of the yields~\cite{duke_fission-fragment_2016,al-adili_fragment-mass_2016,gooden_energy_2016}. Such behavior may explain the difference between the two data sets of Flynn for \Fm256 \cite{flynn_distribution_1972,flynn_distribution_1975}, as well as the high symmetric yields obtained in \Fm254 compared with spontaneous fission experiments. For \Fm258, the situation is the opposite: increasing the energy is expected to flatten the main symmetric peak. This is actually the behavior that we obtain when increasing the energy of the initial state of our dynamic calculation (see Sec.~\ref{sec:ei}). As the energy increases, the wave packet spreads more easily and populates the modes that are not the most energetically favorable. This is consistent with the fact that the experimental data by Hoffman and Hulet associated with spontaneous fission are much more peaked compared with the data of Flynn and our results for induced fission.

\begin{figure}[!htb]
\includegraphics[width=0.45\textwidth]{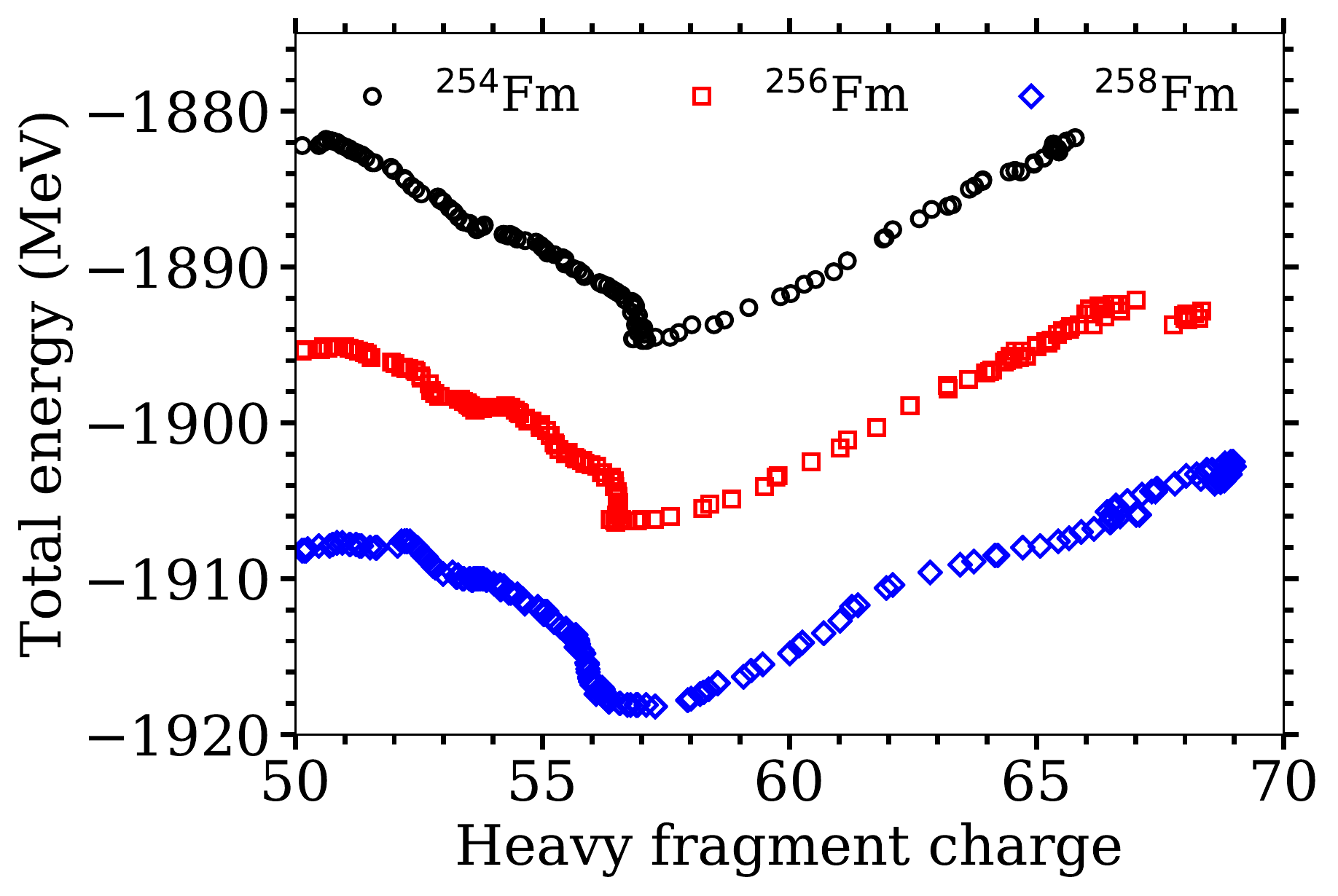}
\caption{Total energy as a function of the heavy fragment charge along the isoline $Q_{\rm N}=7.5$.}
\label{fig:front}
\end{figure}

One should emphasize that changes in the principal fission modes can not be detected when looking only at the structure of scission configurations along the frontier where the flux is computed.
At first glance, this seems totally inconsistent with the fact that scission-point models such as Ref.~\cite{lemaitre_new_2015,pasca_charge_2018} could be able to reproduce this transition between symmetric and asymmetric yields for Fermium isotopes. 
In such models, the statistical population of a given mass and charge split is often given by a Boltzman factor that depends on the free energy at the scission configurations of interest. 
Fig.~\ref{fig:front} shows that the collective potential energy as a function of the proton number of the heavy fragment is remarkably similar for all three isotopes (notwithstanding a trivial shift due to the binding energy of the extra neutrons). A thermal occupation of these 'scission' configurations (along the $Q_N=7.5$ isoline) would be very similar and should result in mostly asymmetric yields for the three Fermium isotopes. 
The fact that statistical models are somewhat capable to reproduce the experimental transition therefore suggests that the scission configurations they are using are rather different from the ones we observe along the $Q_N=7.5$ isoline. More precisely, we should expect that these configurations correspond to geometrical shapes that are somewhat equivalent to the shapes we observe in our calculations in the area around $Q_{20}=180$ b, where the system ``chooses'' between the two different modes.

\subsection{Structure of Asymmetric Modes in \Fm254}
\label{sec:two_modes}

\begin{figure}[!ht]
\includegraphics[width=0.45\textwidth]{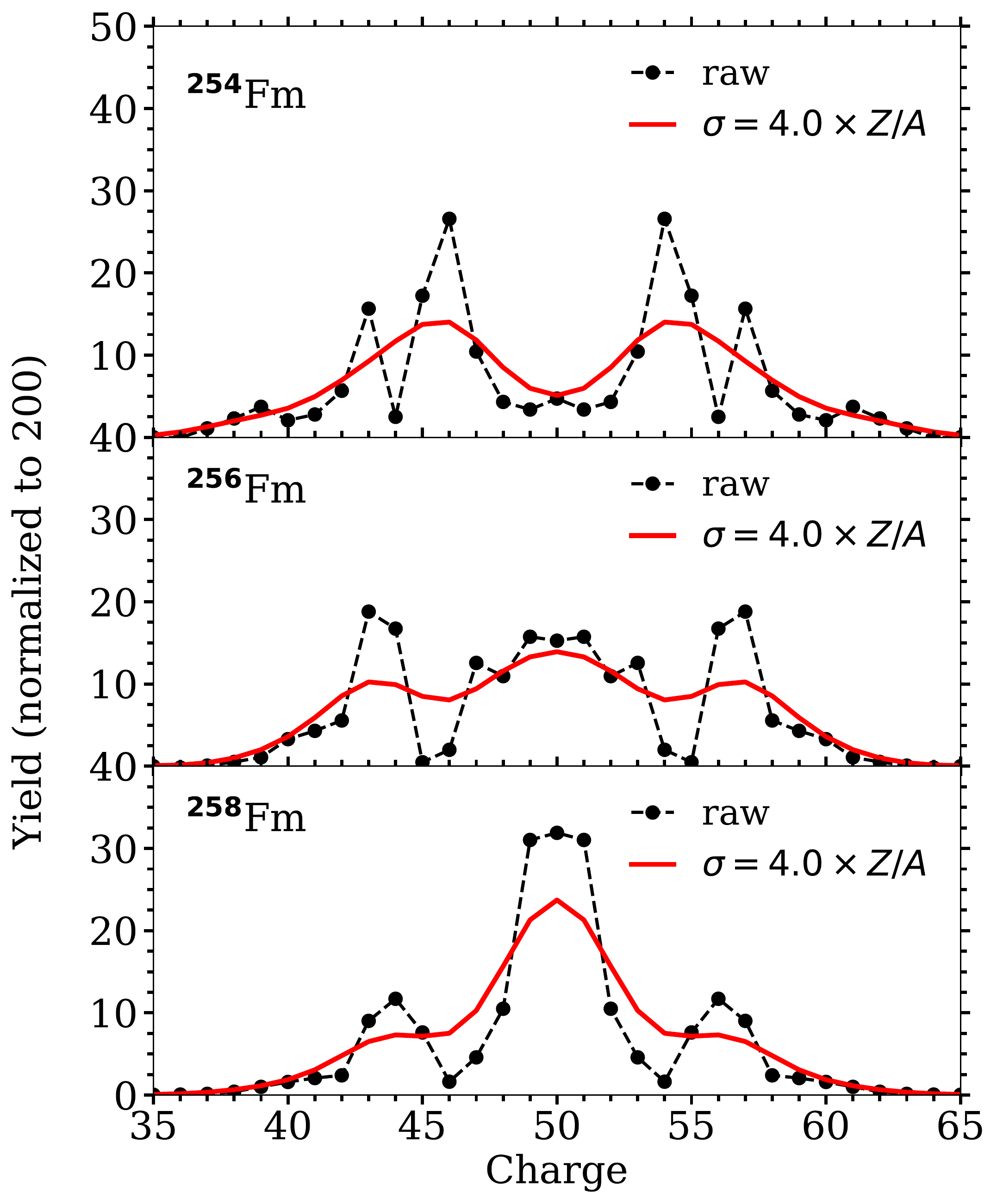}
\caption{Primary fragment charge yields (normalized to 200\%) obtained with the Gogny {\DONES} effective interaction. The black doted line represents raw results directly obtained from the flux through the frontier, whereas the red full line accounts for the convolution of the raw results with a Gaussian of width $\sigma=4.0\times Z/A$.}
\label{fig:fmtrans_charge}
\end{figure}

Looking more closely at the fission of \Fm254, we found that the large asymmetric peak in the mass yields is actually coming from two well-separated valleys. This is particularly visible in Fig.~\ref{fig:fmtrans_charge}, where we show the charge yields obtained without any convolution with a Gaussian form factor. One asymmetric mode is centered at $Z=57$ while the other one lies around $Z\simeq 54$. The first one corresponds to the output of the large asymmetric valley corresponding to configurations  around $(350, 50)$ in barn units. It corresponds to rather elongated configurations. 
The other one corresponds to a tiny valley starting at lower elongation and asymmetry, around $(270,25)$ barn units.
Looking at the evolution of the raw yields as a function of neutron number, we find that the most asymmetric mode in the Fermium chain (the one centered on $Z=57$ for the heavy fragment) is pretty stable. When the number of neutron increases, it is the less asymmetric mode that vanishes and becomes the symmetric mode. We may speculate that these two asymmetric modes could well be related to the standard-1 and standard-2 modes widely used to fit actinides fission yields~\cite{brosa_nuclear_1990}.

\section{Stability of the Results}
\label{sec:discussion}

\subsection{Parametrizations of the Gogny EDF}

The main input of the {\TDGCM}+{\GOA} approach for the determination of fission yields is the energy density functional underpinning all the calculations. Although this input should ultimately be related to the bare interaction between nucleons, practical applications in heavy nuclei rely on empirical parameterizations fitted on various key nuclear observables. In the case of the Gogny effective interaction, the three major parameterizations differ in the methods adopted for the fitting procedure. Although it is the oldest one, the {\DONES} parameterization~\cite{decharge_hartree-fock-bogolyubov_1980,berger_time-dependent_1991} includes constraints on the fission barrier height of \Pu240 estimated at the {\HFB} level and can give a rather good description of most nuclear properties. For this reason, we have used it in this work as a reference. The {\DONEN} parameterization~\cite{chappert_towards_2008} was designed to better reproduce the properties of neutron matter at the {\HFB} level and is therefore expected to perform better in the neutron-rich sector of the nuclear chart. Finally, {\DONEM}~\cite{goriely_first_2009} was especially designed to reproduce the masses and radii of the entire nuclear chart at the 5DCH level, i.e. within a static {\GCM}+{\GOA} framework including all quadrupole degrees of freedom. 
The impact of the choice of parameterization of the Gogny interaction on some fission properties such as barrier heights and half-lives has been investigated in Ref.~\cite{rodriguez-guzman_microscopic_2014}. Although it is clear that significant differences appears between parameterizations, e.g. {\DONES} underestimates nuclear binding energies compared with the two others, the topology of the least-action fission paths are qualitatively similar. Therefore, the impact on the fission yields can only be tested in a fully dynamical calculation.

\begin{figure}[!ht]
\includegraphics[width=0.45\textwidth]{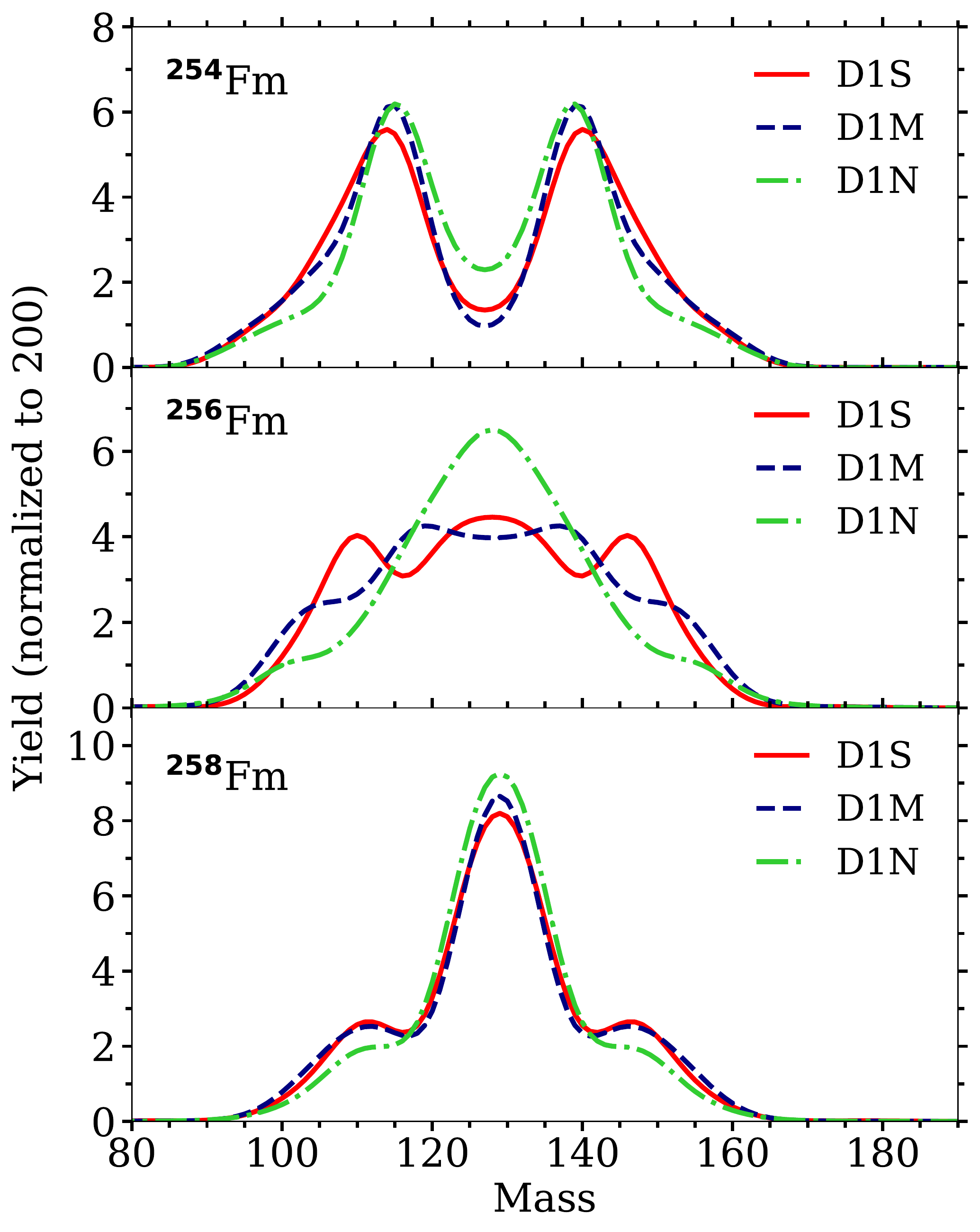}
\caption{Comparison of the primary fragment mass yields obtained with the {\DONES}, {\DONEN} and {\DONEM} parameterizations of the Gogny force. All the yields are normalized to 200\%.}
\label{fig:fmtrans_inter}
\end{figure}

In this section, we compare the fission yields obtained from the three parameterizations. All the codes and numerical parameters are exactly the same for each calculation, which provides for the first time a clean view of the sensitivity of the yields to the Gogny parameterization. The results are plotted in Fig.~\ref{fig:fmtrans_inter}.
%
%
The most important conclusion of this study is that the transition from asymmetric to symmetric fission in Fermium isotopes holds for all three interactions. In fact, results from the different parameterizations in \Fm254 and \Fm258, where one of the modes is strongly favored, are remarkably close. This suggests that for this kind of nuclei, the {\TDGCM}+{\GOA} method provides a robust method to predict the qualitative feature of the yields.
On the other hand, the yields obtained for \Fm256 differ significantly. The {\DONEN} effective interaction gives a wide symmetric peak whereas the yields are pretty flat for {\DONES} and {\DONEM}. In this transition nucleus, the sensitivity to the details of the energy functional is much more pronounced. Since the yields result from the competition between several modes, results are much more sensitive to the small changes in the {\PES} topology that different parameterizations can induce. In such nuclei, the {\TDGCM}+{\GOA} is much less predictive, mostly because of our lack of constraints on the underlying {\EDF}. On the other hand, if all other limitations of the {\TDGCM}+{\GOA} could finally be taken care of, these transition nuclei could provide good test benches to validate energy density functionals.

\subsection{Initial state} 
\label{sec:ei}

The goal of this section is twofold: first to study the impact of the initial energy of the fissioning system and try to assess how meaningful the comparison with experimental data shown in Fig.~\ref{fig:theovsexp} is; second to check that changing the Gaussian width $\sigma_i$ used to build the initial state within a reasonable range does not affect our global conclusions.

%
\begin{figure}[!ht]
\includegraphics[width=0.44\textwidth]{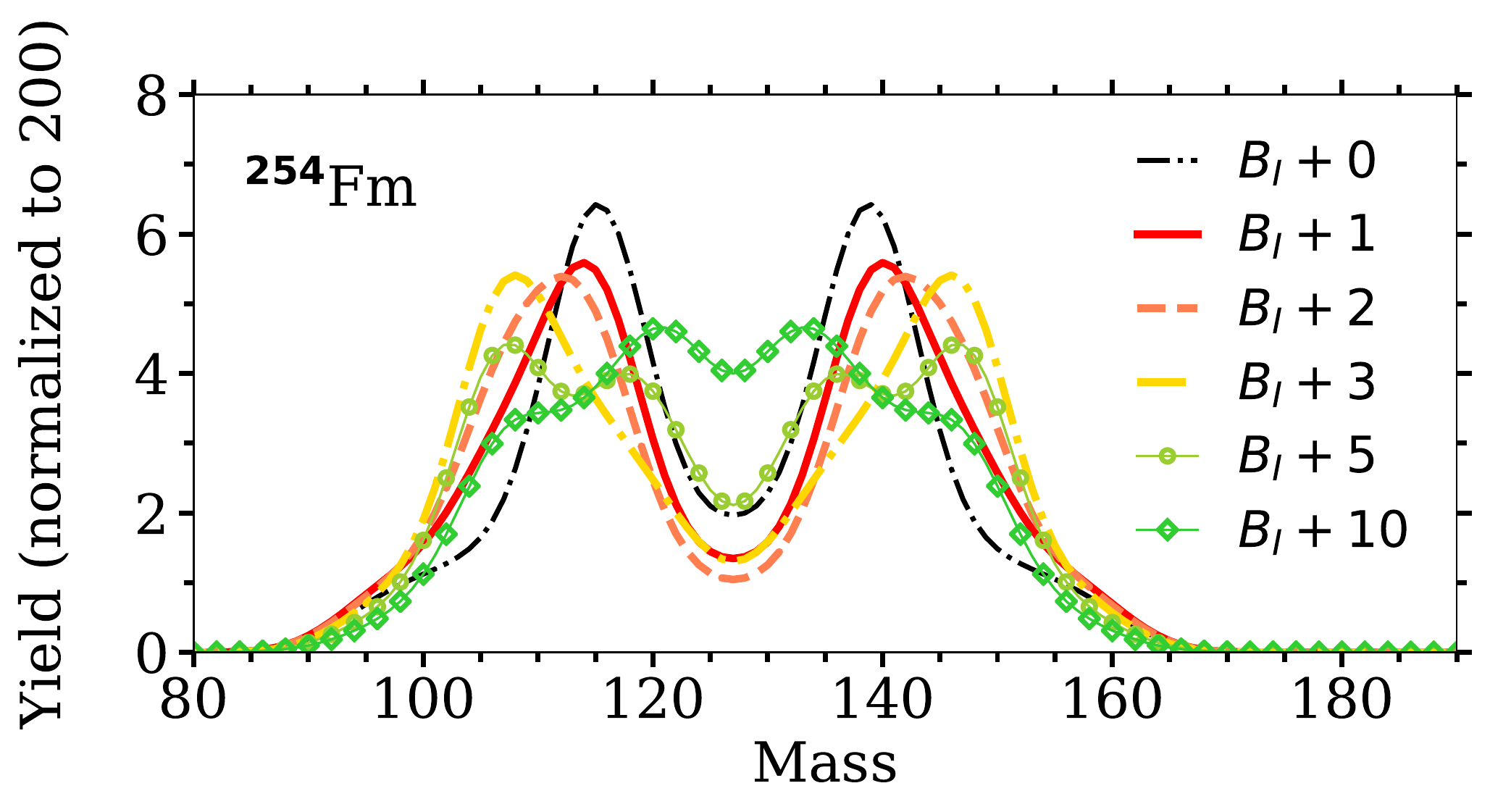}
\includegraphics[width=0.45\textwidth]{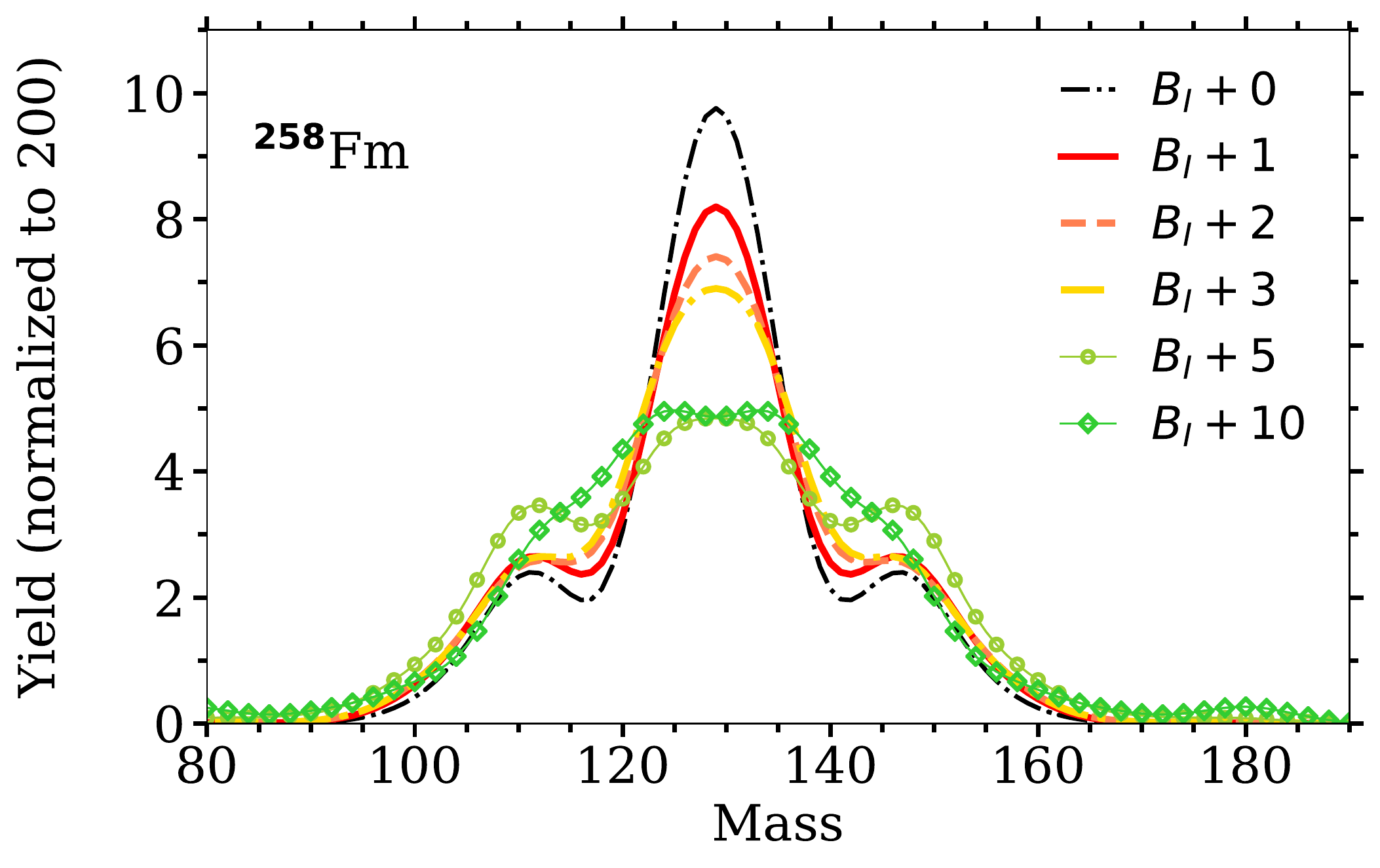}
\caption{Evolution of the primary fragment mass yields as a function of the initial energy for \Fm254 and  \Fm258. The energy is given in MeV relative to the energy of the saddle point of the first fission barrier.}
\label{fig:fm258_ya_ei}
\end{figure}
%
For thermal neutron-induced fission, the initial energy should be the neutron separation energy of the studied Fermium. This is typically $S_{n}\simeq 6$ MeV according to the ENSDF database~\cite{noauthor_evaluated_nodate}. In addition, the neutron-induced fission of \Fm256 and \Fm258 is known to occur already significantly with a thermal neutron beam as reported in Ref.~\cite{flynn_distribution_1975}. This means that the initial energy of the system is higher than the fission barrier, $\Delta E_i = (E^*_i - B_I ) > 0$. Since the fission barrier energy is positive for these systems, it means that the initial energy relative to the fission barrier should be in the range
\begin{equation}
 S_n > \Delta E_i > 0
\end{equation}
This is typically the range of energies that the {\TDGCM}+{\GOA} calculation can probe. To assert the sensitivity of the fission yields in this energy range, we performed a series of calculations with various initial energies. The results are reported in Fig.~\ref{fig:fm258_ya_ei}. 
For \Fm254, the main effect of an increase of the initial energy is a progressive shift of the asymmetric peak toward more asymmetric fragmentations. For the extreme case of $E_i = B_I  +10$ MeV, the fission yields become completely different, which is a consequence of the collective wave packet spreading without being so much influenced by the topology of the {\PES}. In the case of \Fm258, the increase of the collective energy also implies a spreading of the fission yields. This is consistent with the experimental data showing strongly-peaked yields for spontaneous fission and much more smoothed ones for induced fission. 
It is important to emphasize that the major modes predicted do not change when varying the initial energy in a range of a few MeV around the fission barrier.

\begin{figure}[!ht]
\includegraphics[width=0.44\textwidth]{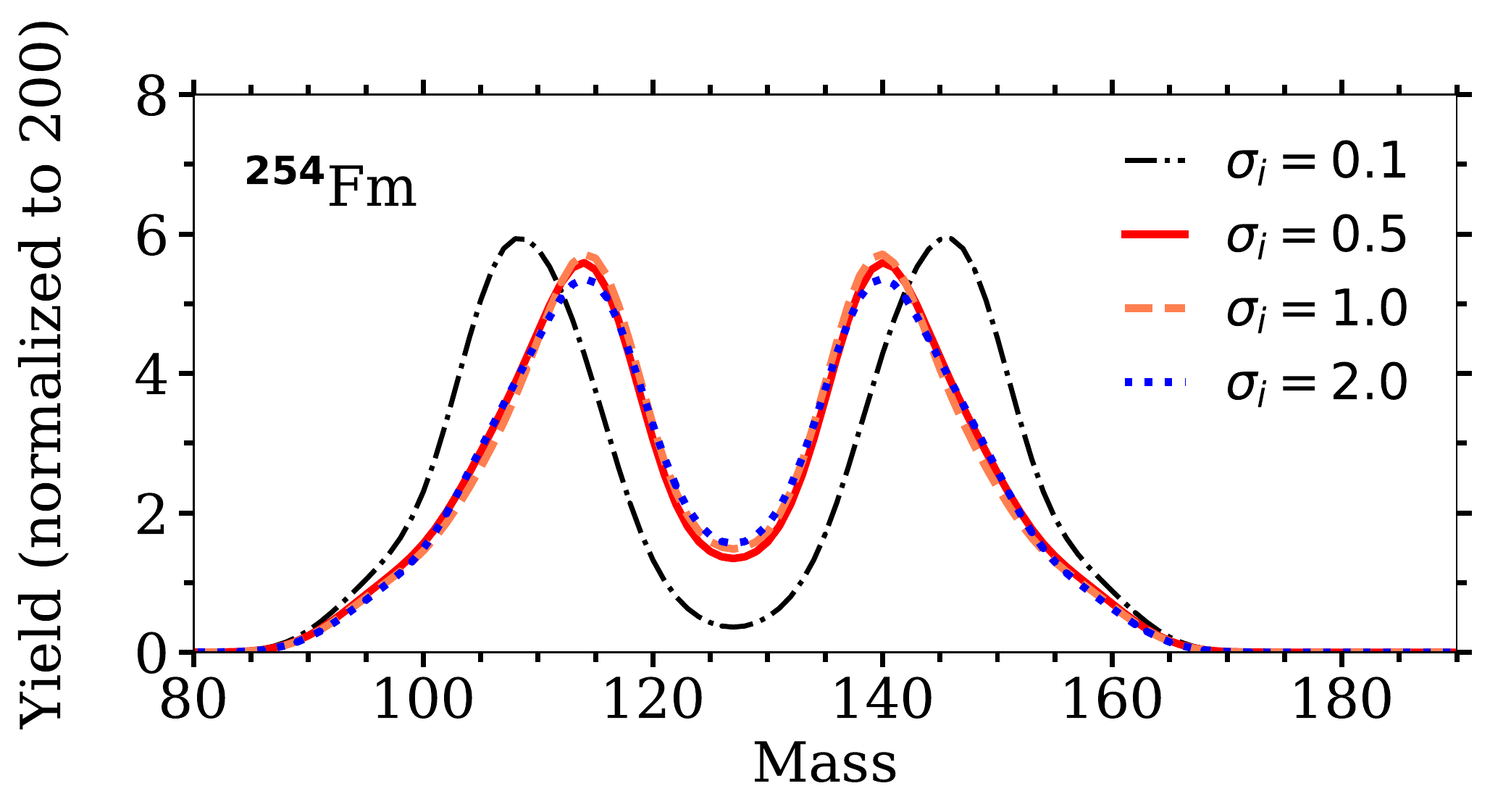}
\includegraphics[width=0.45\textwidth]{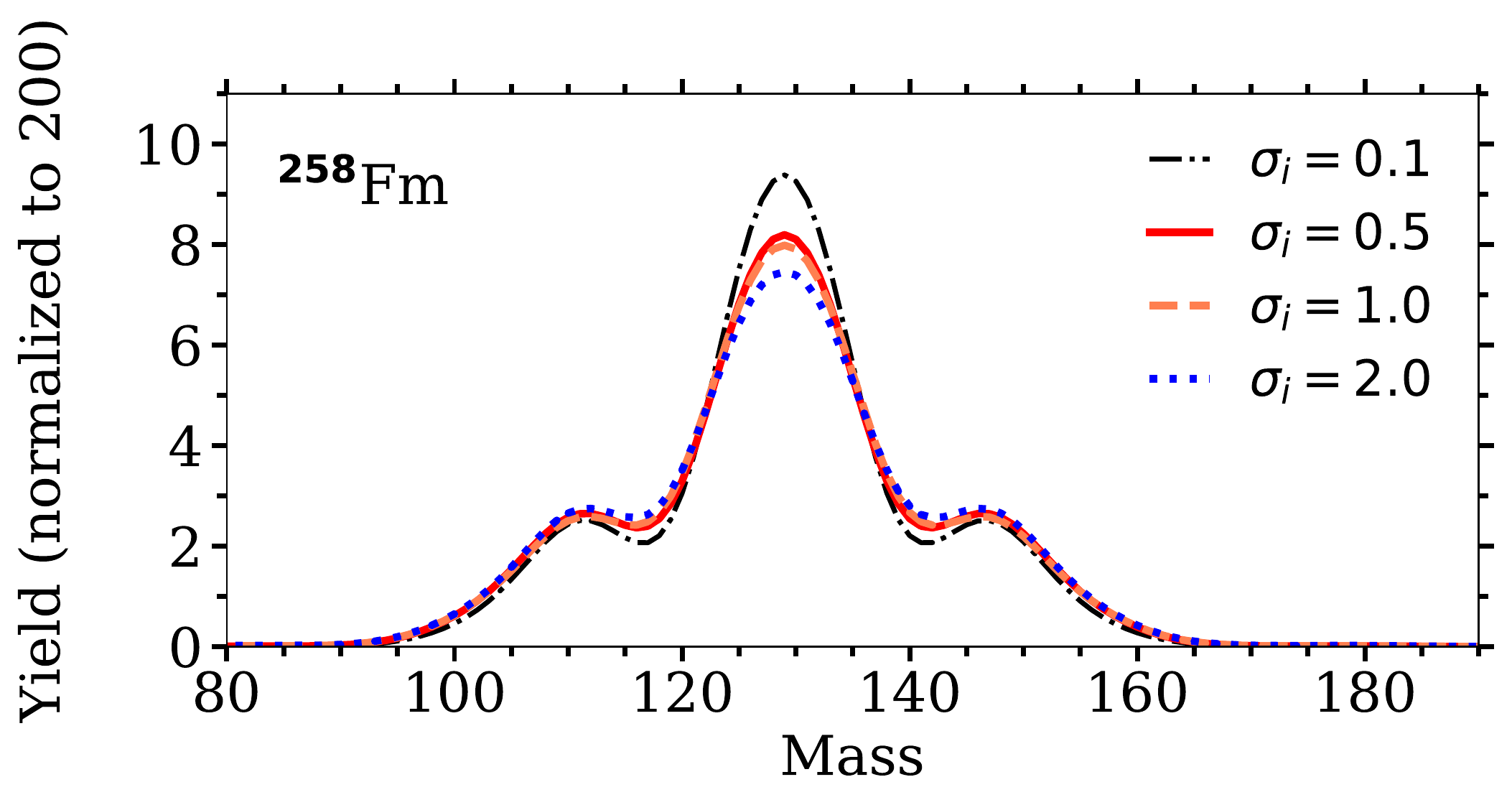}
\caption{Primary fragment mass yields computed with different Gaussian width $\sigma_i$ to build the initial wave packet. The width is given in MeV, and all yields are normalized to 200\%.} 
\label{fig:fm258_ya_isig}
\end{figure}
%

Although the initial energy of the system may be known in fission experiments, the quantum state of the compound system is not. We assumed here that the deformation of the initial state should be close to the one of the ground state with some fluctuation related to its excitation energy. To be conclusive, our results should however not depend too much on the details of the initial state. 
The impact of the choice of the initial state on {\TDGCM}+{\GOA} yields has already been explored in the fission of \Pu240 and \Cf in Ref.~\cite{regnier_fission_2016,zdeb_fission_2017}. In these two cases, the characteristics of the main fission mode were not drastically affected by initializing the dynamics with different types of collective states (boosted Gaussian, Gaussian or Fermi mixing of eigen-modes of an extrapolated first potential well).
To check the robustness of our conclusion in the case of the Fermium isotopes, we performed calculations with different values of the Gaussian width used to build the initial wave packet $\sigma_i = 0.1, 0.5, 1.0$ and $2.0$ MeV. In Fig.~\ref{fig:fm258_ya_isig} we show that the symmetric/asymmetric transition predicted holds whatever the value of $\sigma_i$. The most notable change occurs for $\sigma_i=0.1$ where the initial state reduces to a single eigen-mode of the extrapolated first well. In this extreme (and somewhat unrealistic) case the yields become indeed more sensitive to the characteristics of the selected eigen-mode.

\subsection{Theory of Collective Motion}

It is well known that building the {\GCM} on a basis made of only time-even generator states fails to capture some aspects of the dynamics of the system~\cite{ring_nuclear_2004}. In the special case of translational motion, this leads to underestimating the collective inertia. To mitigate this issue, one possibility is to simulate the collective dynamics of the fissioning nucleus within the requantized adiabatic time dependent Hartree-Fock-Bogoliubov ({\ATDHF}) theory~\cite{baranger_adiabatic_1978}. Indeed, the Pauli requantization scheme yields an evolution equation formally identical to Eq.~\eqref{eq:tdgcmgoa}, where the inertia becomes the {\ATDHF} inertia, the metric is the determinant of this inertia tensor, and the collective potential does not contain any zero-point energy contribution.
%
\begin{figure}[!ht]
\includegraphics[width=0.45\textwidth]{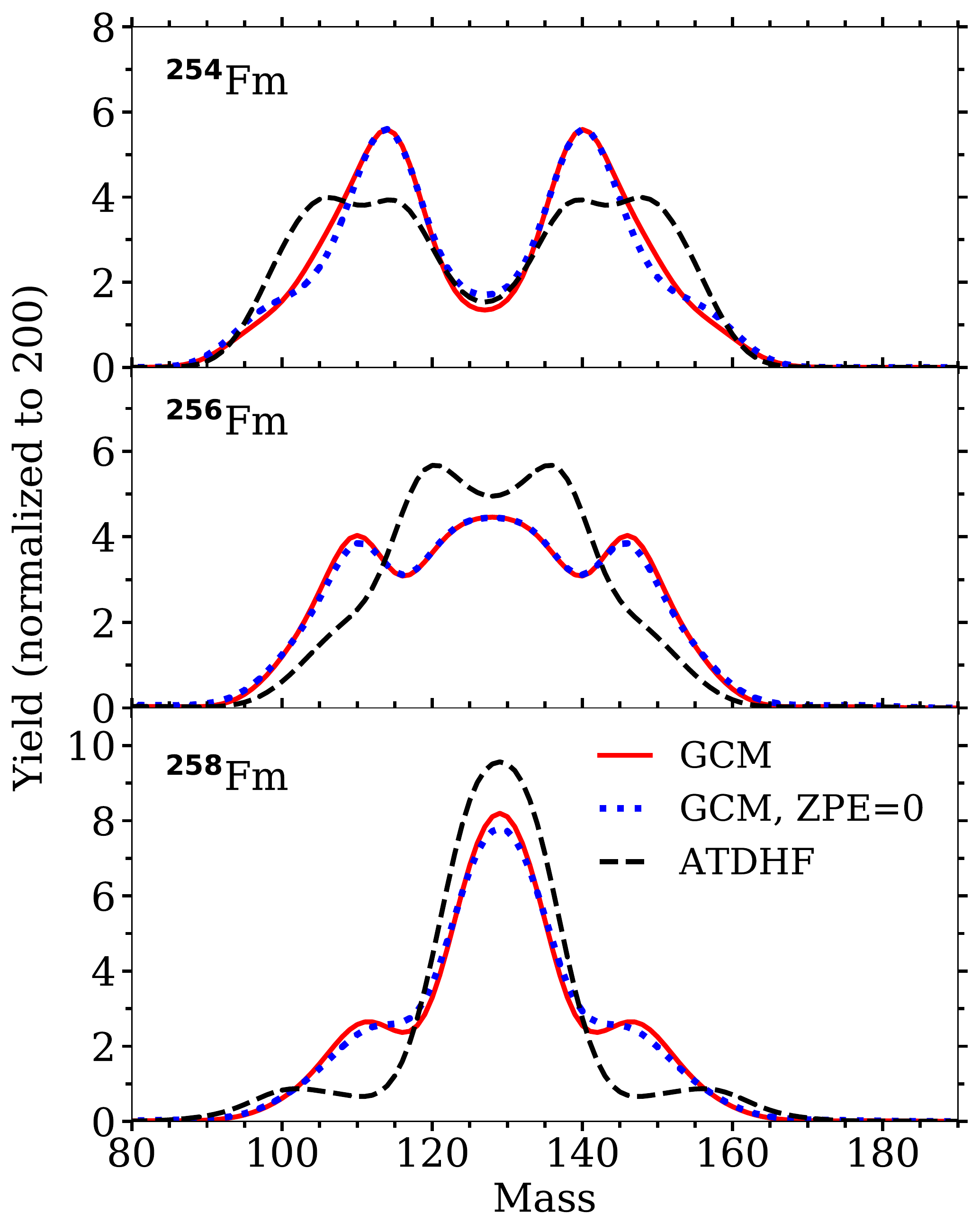}
\caption{Comparison of the primary mass yields obtained with the {\GCM} (full red line) and the {\ATDHF} (dashed black line) collective motion approaches. For completeness, we also give the results obtained with the {\GCM} approach without the zero point energy correction on the potential landscape (doted blue line). All the yields are normalized to 200\%.}
\label{fig:fmtrans_mass}
\end{figure}

Figure~\ref{fig:fmtrans_mass} shows the comparison of primary mass yields in Fermium isotopes between the {\ATDHF} and {\GCM} prescriptions. Although, the fission yields are significantly impacted by this change in the many-body method, the asymmetric/symmetric transition is still predicted at the correct neutron number for both of them.
However, it is clear that the discrepancies resulting from different treatments of the collective dynamics ({\ATDHF} versus {\GCM}) are more pronounced than the ones resulting from different choices of effective interactions, cf. Fig.~\ref{fig:fmtrans_inter}.
To pinpoint even more precisely the origin of these discrepancies, we computed the {\GCM} dynamics without including the zero-point energy correction ({\ZPE}) to the potential; see Fig.~\ref{fig:fmtrans_mass}. Clearly, the energy correction plays a marginal role in determining the fission fragment mass distribution, and the main source of differences between {\ATDHF} and {\GCM} results is the collective mass tensor. 
This suggests that including the physics of time-odd components into the {\GCM} (for instance as proposed in Ref.~\cite{goeke_generator-coordinate-method_1980}) should be a priority if one is to improve the accuracy of these predictions.

%
\subsection{Position of the Frontier}
\label{sec:front}

The definition of the frontier in our approach is strongly constrained by the discontinuity between the fission and fusion valleys. For the {\TDGCM}+{\GOA} to be valid, the dynamics should only take place in a continuous manifold of generator states. As a consequence, the frontier must be before the transition from the fission to the fusion valley.
In this paper we choose to compute the frontier as an isoline of the neck particle operator $\hat{Q}_N$. For this isoline to be before the fusion valley, we find that $Q_{\rm N}$ must be at least greater than 7.0. To simulate the dynamics up to configurations that are as close as possible to scission, the best choice is to put the frontier at the lowest possible isovalues of $Q_{\rm N}$ and we therefore choose $Q_{\rm N}=7.5$. 

As mentioned in Section \ref{subsubsec:yields}, this definition of the frontier implies calculating the yields from configurations that still contain a sizeable number of particles in the neck, which results in a non-negligible nuclear interaction energy between the fragments. This is an intrinsic limitation of our 2-dimensional collective description of the process. Going beyond would require adding some missing intermediate states close to scission into the {\GCM}. This could be achieved either by systematically adding some collective degrees of freedom, at the price of an exponential increase of the numerical cost, or by finding a better manifold of states connecting continuously the fission and fusion valleys. In both cases, such a study is beyond the scope of this work.

\begin{figure}[!ht]
\includegraphics[width=0.45\textwidth]{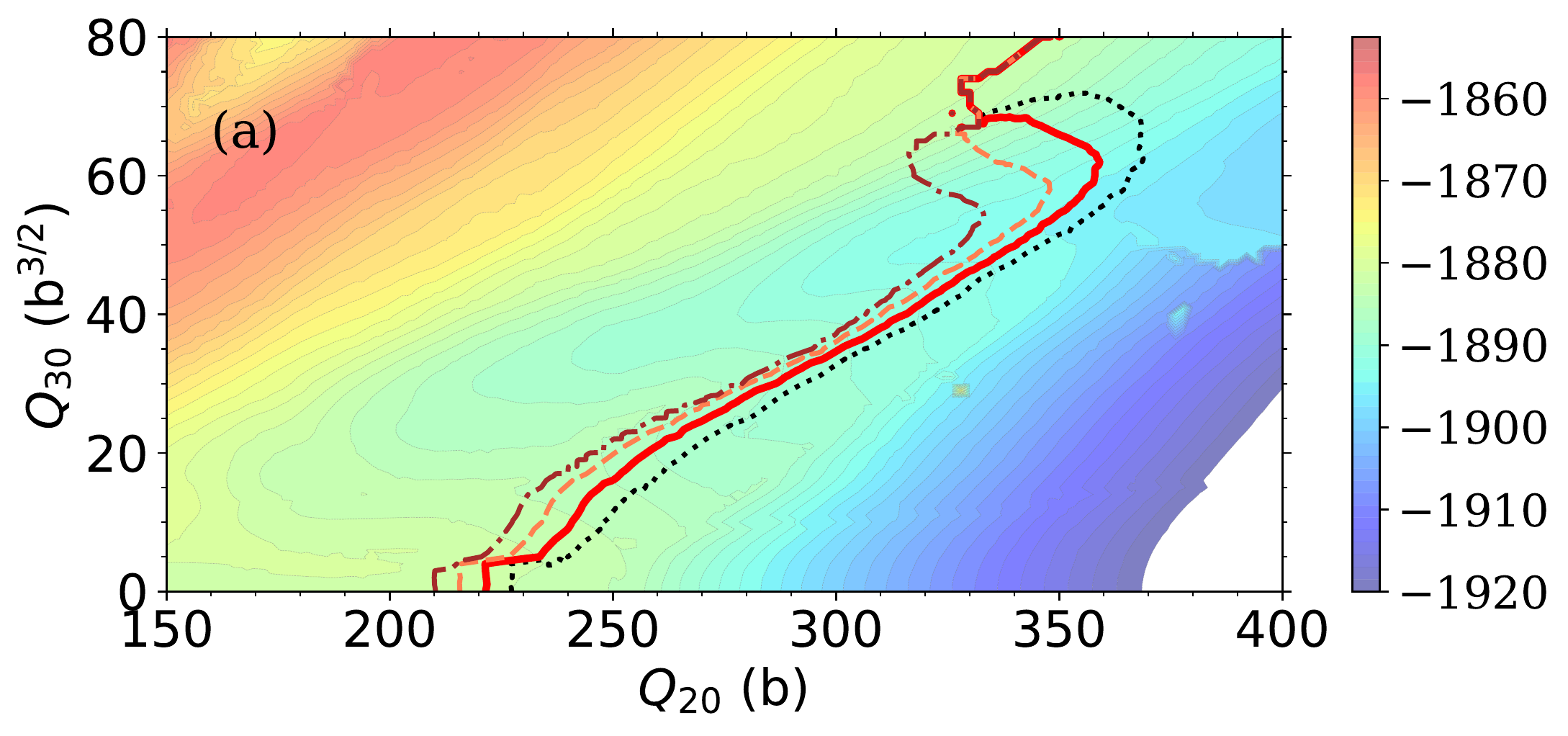}
\includegraphics[width=0.45\textwidth]{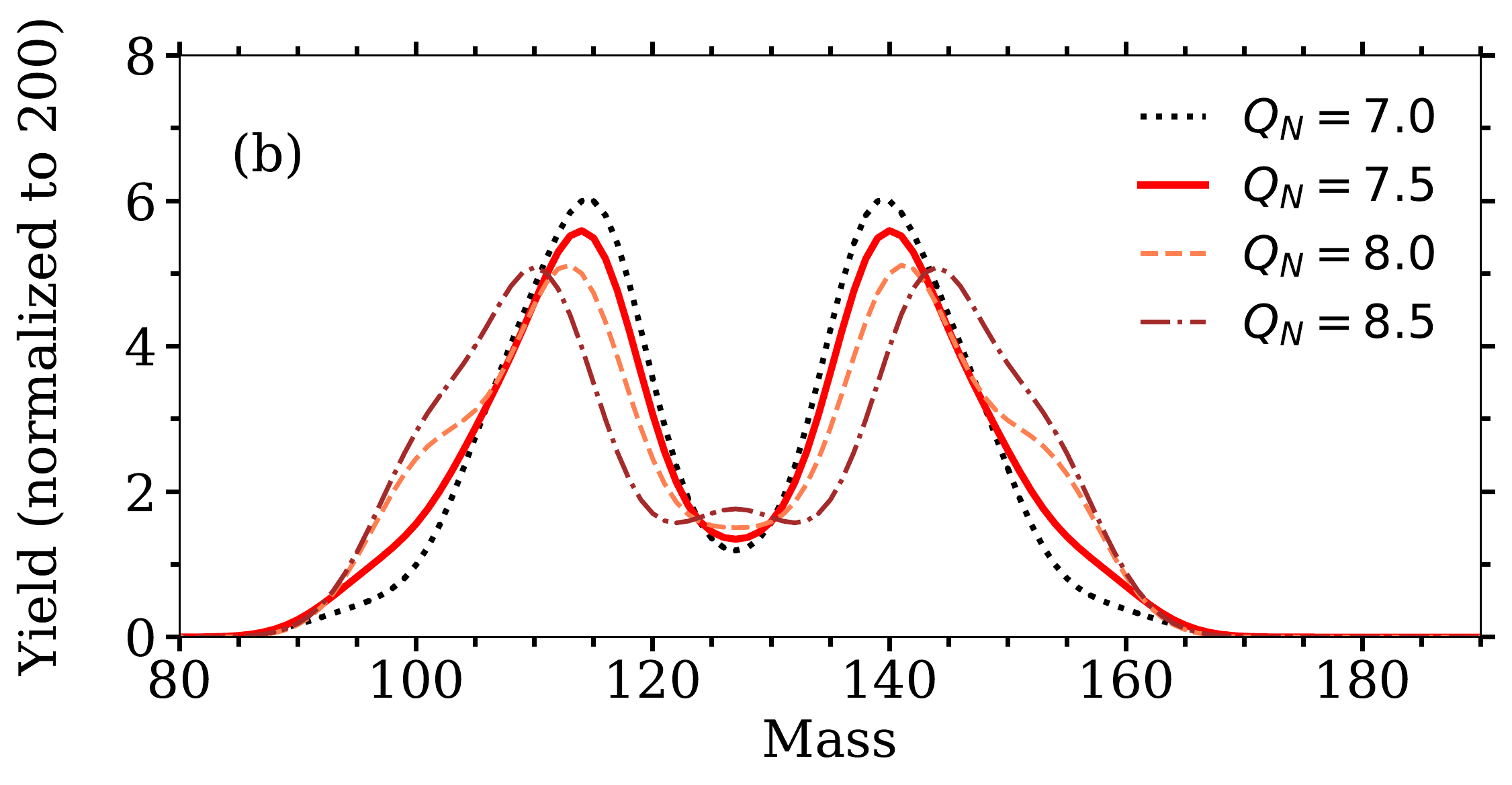}
\caption{(a): Isolines $Q_{\rm N}=7.0,7.5,8.0,8.5$ of the Gaussian neck operator used as frontiers to compute the fission yields of \Fm254. (b): Variation of the primary fragment mass yields of \Fm254 with the neck operator isoline used as frontier.}
\label{fig:ya_qn}
\end{figure}

To assess the uncertainty coming from the arbitrary position of the frontier, we computed the yields for different frontiers defined by $Q_{\rm N}=7.0, 7.5, 8.0, 8.5$. Figure~\ref{fig:ya_qn} shows the location of these frontiers on the {\PES} as well as the yields obtained for the case of \Fm254.
%
%
Although the modification of the frontier does impact the details of the resulting yields, the asymmetric fission picture remains unchanged. This is consistent with the fact that the dominant fission mode is determined at rather low quadrupole deformations ($Q_{20} \simeq 180$ b), much before reaching these frontiers. 
Similar results were found for the fission of \Fm256 and \Fm258.
One might use the results of Fig.\ref{fig:ya_qn} to estimate the yields at the asymptotic limit of vanishing values of the neck.

\subsection{Discontinuities in the $(Q_{20},Q_{30})$ Manifold} 
\label{subsec:discont}

To be mathematically valid, the {\TDGCM}+{\GOA} formalism requires a continuous and twice differentiable manifold of generator states. In practice, the {\PES} obtained by series of constrained {\HFB} calculations does not necessarily satisfy this property. As stated in Ref.~\cite{dubray_numerical_2012} a {\PES} may contain discontinuous {\HFB} states. 
To detect the presence of such discontinuities, we need to define a distance between {\HFB} states. A fully quantum-mechanical distance could be provided based on the calculation of the overlaps between any pair of states~\cite{verriere_fission_2017,verriere_description_2017}. However, in this paper we use a much simpler metric $D$ based on the one-body local density,
\begin{equation}
D(\qvec,\qvec') = \int |\rho_{\qvec}(\rvec) - \rho_{\qvec'}(\rvec)| d^3\rvec ,
\end{equation}
where $\qvec$ and $\qvec'$ refer to two {\HFB} states of the {\PES} and $\rho_{\qvec}(\rvec)$ and $\rho_{\qvec'}(\rvec)$ are their respective local, one-body local densities.
This distance is only sensitive to the diagonal one body-density and does not involve the anomalous density. As a consequence, this metric may miss some discontinuities, in particular the ones related to pairing correlations.

\begin{figure}[!ht]
\includegraphics[width=0.5\textwidth]{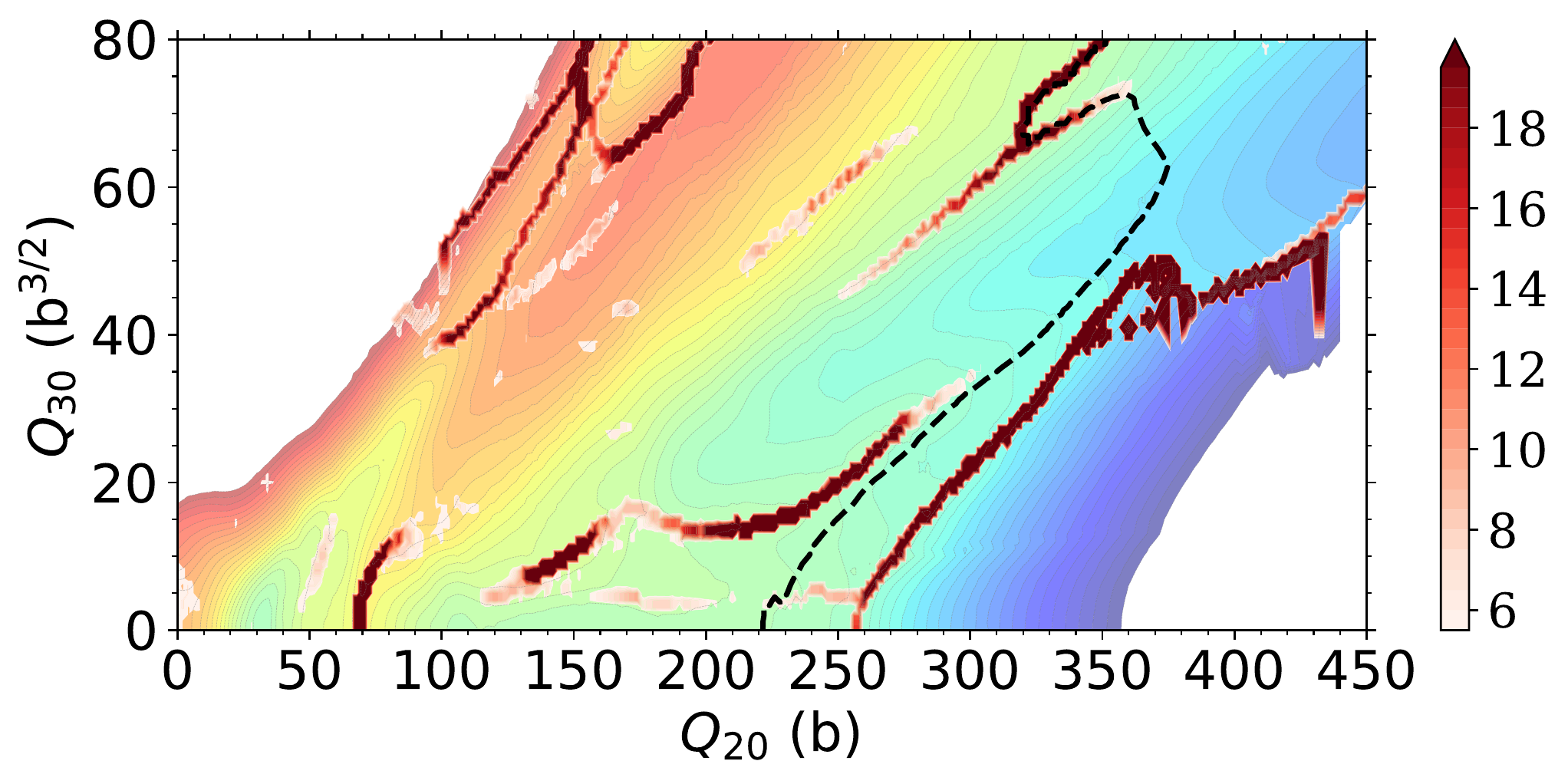}
\caption{Discontinuity indicator plotted on top of the \Fm256 {\PES}. The red color scale represents the value of the discontinuity indicator $I(k)$. Values below 5.5 are not plotted. The background color map represents the potential energy surface. Finally, the black dashed line is the frontier corresponding to $Q_{\rm N}=7.5$. For the sake of legibility, we removed the points belonging to the non-converged island of high energies in the fusion valley (see Fig.~\ref{fig:fmtrans_pes} for comparison).}
\label{fig:fm256_dist}
\end{figure}

To check the quality and validity of our 2-dimensional {\PES}, we compute for each {\HFB} state $\qvec$ the discontinuity indicator $I(\qvec)$,
\begin{equation}
I(\qvec) = \text{max} \left\{ D(\qvec,\qvec') \, |\,  \forall\,  \qvec' \text{neighbor of }\qvec \right\} .
\end{equation}
We show in Fig.~\ref{fig:fm256_dist} how this indicator allows us to identify discontinuities between neighboring areas of the potential energy surface for \Fm256.
%
%
By plotting the highest values of the discontinuity indicator, we clearly see various lines which correspond to sharp discontinuities between neighboring {\HFB} states. The topology and geometry of these lines were found to be similar for the three Fermium isotopes and they could be classified as follows:

\paragraph{North-west sector of the {\PES}} 
In this region of exotic shapes, the density of discontinuities is rather high. However, the potential energy associated with such configurations is at least 10 MeV above the energy of the {\GCM} ground state and quite far from the principal fission valleys. The collective wave packet does not populate this area during the evolution and the associated discontinuities therefore have no impact on the resulting yields.

\paragraph{Top of the first potential barrier}
A discontinuity line is present at the top of the first fission barrier at elongations around $Q_{20}=70$ b. This typically indicates that our 2-dimensional description underestimates the height of this first potential barrier. While this could impact significantly the calculation of fission half-lives for example, we expect that it is less important for the mass distributions, since this discontinuity does not affect the competition between the different fission modes. To gain some information on the potential influence of this discontinuity on the fission yields, a 3-dimensional study involving the hexadecapole mass moment operator $\hat{Q}_{40}$ should be done in the future.

\paragraph{Ridge between the main symmetric and asymmetric valleys}
Between the two main asymmetric and symmetric valleys also lies a discontinuity line at the top of the potential ridge. The values of  $Q_{40}$ are lower in the symmetric valley than in the asymmetric valleys. At the frontier between the symmetric and asymmetric valleys, a gap in $Q_{40}$ values can be seen. This is a signature of two separated valleys in the 3-dimensional space $(Q_{20},Q_{30},Q_{40})$ that are now overlapping in our 2-dimensional working space. However, this discontinuity line is roughly parallel to the direction of the main flow of the collective wave packet, which follows the bottom of the valleys. 
In the case of \Fm254 and \Fm258, the collective wave packet follows mostly one valley and we expect that the spurious flux crossing this line is small compared to the total flux crossing the frontier. 
In this scenario, the fission yields would not be so much impacted. 
On the other hand, for \Fm256, this discontinuity could drastically affect the competition between the symmetric and asymmetric modes. This could be partly responsible for the strong symmetric component found in \Fm256 for which the potential energy in this region is pretty flat.

\paragraph{Fission/fusion transition}
Finally, a discontinuity line starting around $Q_{20}=250$ b for symmetric configurations and going up to large asymmetries corresponds to the transition from the fission valley to split configurations. This 'scission' discontinuity has already been extensively discussed in the literature (see for instance Ref.~\cite{dubray_structure_2008,younes_microscopic_2009}). It is one of the main limitations of our approach as it imposes to select a frontier on its left hand side and therefore compute the yields on a set of configurations with a high neck operator value.

To conclude on the subject of discontinuities, we clearly see that they are present in the fission valleys in our 2-dimensional description. With the exception of the 'scission' one, these discontinuities are mostly signaled by a jump in the value of the $Q_{40}$ multipole moment. Adding this variable into our dynamical description would therefore remove most of these ``internal'' discontinuities.

\section{Conclusion}

We computed the primary fragment mass and charge yields for the low-energy induced fission of $^{254,256,258}$Fm within the {\TDGCM} under the Gaussian overlap approximation. The results obtained with the {\DONES} parameterization of the Gogny effective interaction successfully reproduce the expected transition from a mostly asymmetric fission for \Fm254 to a mostly symmetric one for \Fm258. This transition is interpreted in the framework of collective dynamics as a competition between different modes that depends on the number of neutrons in the system. Most of the physics of the transition can already be inferred from the static analysis of the {\PES} and we show that the bifurcation point responsible for the transition happens at quite low elongations $Q_{20}\simeq 180$ b. In addition, our calculations suggest two asymmetric modes for the fission of \Fm254.
The sensitivity of our results to all the inputs of the calculation has been tested and we find that the qualitative picture is robust. Finally, we show that one of the main limitations of this approach is the presence of discontinuities that appear even at low deformation inside the fission valley. In the case of the Fermium isotopes considered in this study, most of these discontinuities are signaled by an abrupt change in the $Q_{40}$ multipole moment value. Extending the calculations to 3-dimensional collective spaces may be sufficient to solve this problem.

\section{Acknowledgements}
We would like to thank H. Pasca for fruitful discussions about statistical fission models.
Support for this work was partly provided through the Scientific Discovery 
through Advanced Computing (SciDAC) program funded by U.S. Department of 
Energy, Office of Science, Advanced Scientific Computing Research and Nuclear 
Physics. It was partly performed under the auspices of the US Department of  
Energy by the Lawrence Livermore National Laboratory under Contract 
DE-AC52-07NA27344. Computing support for this work came from the Lawrence 
Livermore National Laboratory (LLNL) Institutional Computing Grand Challenge 
program.

\appendix

\section{Spectral element discretization of the collective dynamics}
\label{ap:setup_param}

The first step in the numerical resolution of the collective dynamics consists in building the spectral element basis spanning the collective space of interest, and expanding the collective Hamiltonian on this basis. To do so, we used a slightly modified version of the tool {\tt flx-setup} provided with the {\FELIX} package. This tool proceeds through several steps to transform the information contained in a raw ensemble of constrained {\HFB} generator states into relevant inputs for the dynamics. For the sake of reproducibility, we summarize in this section the main steps of this setup and report in Tab.~\ref{tab:setup_param} the exhaustive  inputs to {\tt flx-setup}. The full details on this setup procedure can be found in Ref.~\cite{regnier_felix-2.0:_2018}.

Starting from a ensemble of $Q_{30}>0$ configurations, the setup tool first select only states having a neck operator value above a certain threshold.  
In order to keep only the fission valley and avoid the discontinuity between the fission and fusion valleys we choose the criterion $Q_{\rm N}>7.0$. This choice is discussed in more details in Sec.~\ref{sec:front}.
The deformation domain is then augmented with an absorption band of width 30 in barn units.
As presented in our previous work, this band contains an additional Hamiltonian term to absorb progressively the collective wave packet and avoid reflections on the boundaries of the domain. The absorption is parametrized by an absorption rate $r=100$ zs$^{-1}$ and a characteristic width $w=30$ in barn units (cf.~\cite{regnier_felix-1.0:_2016}).

Guided by the numerical convergence benchmarks performed on \Fm256 in Ref.~\cite{regnier_felix-2.0:_2018}, we choose to discretize the collective Schr\"odinger equation on a spectral element basis built with degree-3 polynomials. The spatial domain is partitioned as a mesh of squared cells of size $h_{20}=4.24$ b, $h_{30}=1.41$ b$^{3/2}$.
Within a distance 50 (in barn units) to the ground-state, we perform one step of h-refinement for those cells for which the energy at the center is lower than 40 MeV above the ground state. This refinement in the first potential well, where the collective wave function has its most rapid variations, accelerates the numerical convergence of the solution with respect to the dimension of the spectral element basis.
Inside the initial domain (defined by $Q_{\rm N}>7.0$), the fields of the collective Hamiltonian are estimated at the nodes of the finite element basis by linear interpolation between constrained {\HFB} results. 
In the absorption band, all the fields are extrapolated continuously based on their distance to the initial domain in the same way as in Ref.~\cite{regnier_fission_2016}.

Once the finite element basis and all the necessary fields are determined in the $Q_{30}>0$ region, the whole domain is symmetrized so that the dynamics is performed in a box containing configurations with both positive and negative octupole moments.
The collective Hamiltonian is assumed to be symmetric with respect to the $z \rightarrow -z$ transformation. One can show that this assumption implies the symmetry of the fields involved in the collective Hamiltonian and the anti-symmetry of the non-diagonal elements of the inertia tensor, under the action of this transformation. 
Note that in the {\FELIX}-2.0 release, the {\tt flx-setup} tool assumes for this operation that all the fields are symmetric. We had to modify this behavior here so that the non-diagonal part of the inertia are instead anti-symmetrized during this step. This was the only modification brought to {\tt flx-setup}.

\begin{table}[ht!]
 \begin{tabular}{ll}
 \hline
 Option for {\tt flx-setup} &  Value \\
\hline
abs-rate                    &    10                   \\                           
abs-width                   &    30                   \\
alpha                       &    5                    \\
cell                        &    hcube                \\
deg                         &    3                    \\
eigen-nstates               &    100                  \\
eigen-tol                   &    1e-13                \\
eigen-vmax                  &    50.                 \\
extrapol-width              &    30.                  \\
gs                          &    30.,0.0              \\
gs-extrapol-radius          &    50.                  \\
gs-hrefine-vmax             &    40.                  \\
mesh-step                   &    4.24,1.41            \\
outer-well                  &    100.,0.0             \\
qN-cut                      &    7.0                  \\
quad-h                      &    gaussLegendre        \\
quad-m                      &    gaussLobatto         \\
saddle-vmax                 &    30.                  \\
scale                       &    1.,1.                \\
v-slope                     &    4e-2                 \\
\hline
 \end{tabular}
 \caption{Inputs used for the setup of the dynamics with {\FELIX}-2.0.}
 \label{tab:setup_param}
\end{table}


\addcontentsline{toc}{section}{References}
\newpage
\bibliographystyle{apsrev4-1}
\bibliography{prc_fm.bib}

\end{document}